\documentclass[10pt]{iopart}
\usepackage{iopams}
\usepackage{graphicx}    
\begin{document}
\title[Optical reference geometry and inertial forces in Kerr-de~Sitter spacetimes]{Optical reference geometry and inertial forces in Kerr-de~Sitter spacetimes}

\author{ Ji\v{r}\'{i} Kov\'{a}\v{r}
\footnote[1]{Jiri.Kovar@fpf.slu.cz} and Zden\v{e}k Stuchl\'{i}k
\footnote[2]{Zdenek.Stuchlik@fpf.slu.cz}}

\address{Institute of Physics, Faculty of Philosophy and Science, Silesian University in Opava, Bezru\v{c}ovo n\'{a}m. 13, 746 01 Opava, Czech Republic}

\begin{abstract}
Optical reference geometry and related concept of inertial forces are investigated in Kerr-de Sitter spacetimes.
Properties of the inertial forces are summarized and their typical behaviour is illustrated. 
The  intuitive 'Newtonian' application of the forces in the relativistic dynamics is demonstrated in the case of the test particle circular motion, static equilibrium positions and perfect fluid toroidal configurations. 
Features of the optical geometry are illustrated by the embedding diagrams of its equatorial plane. The embedding diagrams do not cover whole the stationary regions of the spacetimes, therefore the limits of embeddability are established. 
A shape of the embedding diagrams is related to the behaviour of the centrifugal force and it is  characterized by the number of turning points of the diagrams. Discussion of the number of embeddable photon circular orbits is also included and the typical embedding diagrams are constructed.   
The Kerr-de Sitter spacetimes are classified according to the properties of the inertial forces and embedding diagrams.   
\end{abstract}


\section{Introduction}
In 1988, a new approach to the relativistic dynamics appeared. By an appropriate conformal (3+1) splitting of the Schwarzschild spacetime, the so-called optical reference geometry was defined
\cite{Abr-Car-Las:1988:GENRG2:}. It was demonstrated that this geometry enables us to define the inertial forces  in the framework of general relativity, providing a description of test particle motion in accordance with our 'Newtonian' intuition \cite{Abr-Pra:1990:RASMN:}, \cite{Abr-Mil:1990:RASMN:}. Later, the definition of the optical reference geometry and related inertial forces was generalized, extending its applicability to any spacetime \cite{Abr-Nur-Wex:1993:CLAQG:}, \cite{Abr-Nur-Wex:1995:CLAQG:}, \cite{Son-Mas:1996:CLAQG:}, \cite{Jon:2006:CLAQG:}. The optical reference geometry was also thoroughly studied in particular spacetimes, such as Schwarzschild-de Sitter \cite{Stu:1990:BULAI:}, \cite{Stu-Hle:1999:PHYSR4:}, Reissner-Nordstr\"{o}m \cite{Kri-Son-Abr:1998:GENRG2:}, Reissner-Nordstr\"{o}m-de Sitter \cite{Stu-Hle:2002:ACTPS2:}, Kerr \cite{Iye-Pra:1993:CLAQG:}, \cite{Stu-Hle:1999:ACTPS2:} or Kerr-Newman \cite{Stu-Hle-Jur:2000:CLAQG:}, illustrated by embedding diagrams \cite{Stu-Hle:1999:CLAQG:}, \cite{Hle:2002:GraPrague:}, \cite{Hle:2001:RAGtime2and3:}, and the inertial forces formalism was applied for solving specific problems in relativistic dynamics \cite{Agu-etal:1996:CLAQG:}, \cite{Nay-Vis:1996:CLAQG:}. Behaviour of the centrifugal force is closely related to the shape of embedding diagrams of the optical geometry, therefore many properties of the relativistic dynamics in the spacetimes were effectively illustrated and directly visualized.     

In the present work we extend our previous studies \cite{Stu-Hle:1999:PHYSR4:}, \cite{Stu-Hle:1999:ACTPS2:} to the case of the stationary and axially symmetric Kerr-de Sitter (KdS) spacetimes, reflecting thus their basic properties incorporating a combined influence of the rotation of source and the cosmic repulsion, recently indicated by wide range of cosmological tests. It is worth to stress that the KdS black-hole spacetimes could be important in understanding astrophysical phenomena exposed around supermassive black holes in giant active galactic nuclei as demonstrated in \cite{Sla-Stu:2005:CLAQG:}, \cite{Stu:2005:MPLA:}.
We show that the application of the inertial forces concept proves to be very effective and provides simpler and intuitive approach to some relativistic problems, as compared to the standard general relativistic methods. 
%---------------------------------------------
The standard methods based on the covariant formulation of physical laws in a given background (geodesic equation, energy momentum conservation, etc.) can be then replaced by the well developed methods familiar from Newtonian physics when inertial forces related to the optical geometry are used, making a significant technical simplification in treating physical processes in curved backgrounds and enabling their interpretation in the Newtonian way.
%--------------------------------------------------------------
Thus the present paper could be also considered to be an introduction in developing a new approach to study astrophysical phenomena. We give some explicit examples how the analysis of the inertial forces could immediately and in a very simple way make an enlightenment of the properties of the circular motion of test particles and perfect fluid.

The structure of the paper is similar to the previous study 
of the optical geometry of the Kerr-Newman spacetimes \cite{Stu-Hle-Jur:2000:CLAQG:}, whereas the content is more extensive, mainly in the parts concerning the inertial forces application in the relativistic dynamics.
\mbox{In section 2}, we briefly present basic general definitions of the optical geometry and inertial forces, focusing on the stationary and axially symmetric spacetimes and circular motion.  
\mbox{In section 3}, KdS geometry is introduced. Although the properties of the KdS spacetimes are well known, we briefly discuss them here again, in order to simply demonstrate the so-called 'Chinese boxes' method, used later in much more complicated cases for discussions of the properties of the inertial forces and embedding diagrams. 
\mbox{In section 4}, the inertial forces in the KdS spacetimes are determined for general circular motion, and their properties are briefly discussed and thoroughly represented for the equatorial circular motion. \mbox{In section 5}, some examples of the relativistic dynamics are shown from the point of view of the inertial forces formalism.  
There are many interesting relativistic problems, which can be simply investigated by using the inertial forces formalism. We present such problems that have been studied by the standard general relativistic methods as well, in order to show immediately the effectiveness and simplicity of the inertial forces approach. Namely the equatorial circular motion is studied, simply providing important findings for the test particle and photon circular motion \cite{Stu-Sla:2004:PHYSR4:}, equatorial static radius and static equilibrium positions of test particles on the axis of symmetry \cite{Stu-Kov:2006:CLAQG:}. Another astrophysically relevant application of the inertial forces formalism is demonstrated in deducing  projection of the relativistic Euler equation that allows determination of the perfect fluid equilibrium toroidal configurations \cite{Sla-Stu:2005:CLAQG:}. \mbox{In section 6}, the embedding diagrams of the equatorial plane of the optical reference geometry of KdS spacetimes are discussed and constructed. The whole optical geometry cannot be embeddable, therefore we discuss limits of embeddability. We establish relations between the shape of the embedding diagrams and the behaviour of the centrifugal force, investigating turning points of the diagrams. Finally, the embeddability of photon circular orbits is discussed and typical embedding diagrams are constructed. \mbox{In section 7}, we summarize results of our investigation of the optical reference geometry formalism in the KdS spacetimes, and compare them with results obtained for previously studied backgrounds.       
%*******************************************************************
\section{Fundamental definitions of the formalism}
%------------------------------------------------------------------
\subsection{General case}
In stationary spacetimes described by a metric $g_{ik}$ (with signature +2), the definition of the optical reference geometry requires an introduction of family of special observers with a timelike, unit, and hypersurface orthogonal \mbox{4-velocity} field $n^i$ and with its \mbox{4-acceleration} field equal to the gradient of a scalar function, i.e., \cite{Abr-Nur-Wex:1995:CLAQG:}
\begin{eqnarray}
\label{1}
n^k n_k=-1,\quad
n^i\nabla_i n_k=\nabla_k\Phi,\quad n_{[i}\nabla_j n_{k]}=0.
\end{eqnarray}
Such a vector field can be chosen in the form 
\begin{eqnarray}
\label{3}
n^i=e^{-\Phi}\iota^i,\quad \Phi=\frac{1}{2}\ln{(-\iota^i\iota_i)},
\end{eqnarray}
whereas it corresponds to the unit \mbox{4-velocity} field of stationary observers parallel to a timelike vector field $\iota^i$. 
Then the local instantaneous \mbox{3-dimensional} space of the observers is described by the metric
\begin{eqnarray}
\label{5}
h_{ik}=g_{ik}+n_i n_k,
\end{eqnarray}
the so-called \textit{directly projected geometry}. Its conformally adjusted metric 
\begin{eqnarray}
\label{7}
\tilde{h}_{ik}=e^{-2\Phi}h_{ik}
\end{eqnarray}
is the so-called \textit{optical reference geometry}.

The \mbox{4-velocity} $u^i$ of a test particle with a rest mass $m$ can be decomposed in the reference frame of the special observers with the \mbox{4-velocity} $n^i$ by using the relation
\begin{eqnarray}
\label{8}
u^i=\gamma(n^i+v\tau^i),
\end{eqnarray}
where $\tau^i$ is the unit spacelike vector parallel to the \mbox{3-velocity} $v^i$ of the particle in the \mbox{3-dimensional} space and $\gamma=(1-v^2)^{-1/2}$ is the Lorentz factor (the speed of light $c=1$). Moreover, there is $\gamma=-n^i u_i$, $\gamma v\tau^i=\gamma v^i=u^k h^i_k$,  where $h^i_k=\delta^i_k+n^i n_k$ is a projection tensor. Note that the vector $\tau^i$ is defined only along the world line of the particle. But in our construction, it is necessary to know how $\tau^i$ changes along $n^i$. This can be done in several different ways (different gauges). The gauge used by M. Abramowicz and his co-workers assures that the Lie derivation of $\tau^i$ with respect to $\iota^i=e^\Phi n^i$ vanishes, i.e., $\mathcal{L}_\iota\tau_k\equiv \iota^i\nabla_i \tau_k-\tau^i\nabla_i\iota_k=0$, which causes that the Coriolis force defined below vanishes in static spacetimes \cite{Abr-Nur-Wex:1995:CLAQG:}. 

The \mbox{4-acceleration} $a_k=u^i\nabla_i u_k$ of the particle can be written in the form 
\begin{eqnarray}
\label{9}
a_k=\gamma^2\nabla_k\Phi +\gamma^2 v(n^i\nabla_i\tau_k+\tau^i\nabla_{i}n_k)+\gamma^2v^2\tau^i\nabla_i\tau_k+(v\gamma\dot{)}\tau_k+\dot{\gamma}n_k,
\end{eqnarray}
where $(v\gamma\dot{)}=u^i\nabla_i(\gamma v)$. By using the spacelike unit vector parallel to $\tau^i$ in the optical reference geometry , i.e., the vector $\tilde{\tau}^{i}=e^{\Phi}\tau^{i}$, its covariant form $\tilde{\tau}_i=e^{-\Phi}\tau_i$, the scalar $\tilde{E}=-\iota^iu_i$, identity $\gamma^2=1+v^2\gamma^2$ and condition of hypersurface orthogonality, the projection of the \mbox{4-acceleration} into the \mbox{3-dimensional} space can be written as
\begin{eqnarray}
\label{11}
a_j^{\perp}=h^k_ja_k=\nabla_j\Phi+(\gamma v)^{2}\tilde{\tau}^i\tilde{\nabla}_{i}\tilde{\tau}_{j}+\gamma^2vX_{j}+\dot{V}\tilde{\tau}_j, 
\end{eqnarray}
where $X_j=n^i(\nabla_i\tau_j-\nabla_j\tau_{i})$ and $\dot{V}=u^i\nabla_i(\tilde{E}v)$.

Now, the inertial force in the \mbox{3-dimensional} space ${F_k'}^{\perp}$, related to the real force by the equation $F_k'^{\perp}=-F_k^{\perp}$, can be decomposed into the sum of the gravitational ${G_k}$, centrifugal ${Z_k}$, Coriolis ${C_k}$ and Euler ${E_k}$ forces familiar from the Newtonian physics, i.e., 
\begin{eqnarray}
\label{12}
{F'_k}^{\perp}=-ma_k
^{\perp}={G_k}+{Z_k}+{C_k}+{E_k},
\end{eqnarray}
where
\begin{eqnarray}
\label{13}
G_k=-m\nabla_k\Phi,\\
\label{14}
Z_k=-m(\gamma v)^2\tilde{\tau}^i\tilde{\nabla}_{i}\tilde{\tau}_k,\\
\label{15}
C_k=-m\gamma^2vX_k,\\
\label{16}
E_k=-m\dot{V}\tilde{\tau}_k.
\end{eqnarray}
Note that in the context of the optical reference geometry approach, the gravitational force ranks among the inertial forces.
%-------------------------------------------------------------------
\subsection{Axially symmetric spacetimes and circular motion} 
In static and spherically symmetric spacetimes, the \mbox{4-velocity} field (\ref{3}) can be chosen with the vector \mbox{field $\iota^i$} being the timelike Killing vector field $\eta^i=\delta^i_t$, which is orthogonal to the spacelike azimuthal Killing vector field $\xi^i=\delta^i_{\phi}$.
However, in stationary and axially symmetric spacetimes, the Killing vector fields $\eta^i$ and $\xi^i$ are not orthogonal in general. Thus the vector \mbox{field $\iota^i$} has to be chosen in order to satisfy the hypersurface orthogonality condition.
It is easy to check that the vector field $\iota^i=\eta^i+\Omega_{LNRF}\xi^i$, where $\Omega_{LNRF}=-\eta^i\xi_i/\xi^i\xi_i$ is hypersurface orthogonal and thus the \mbox{4-velocity} field  
\begin{eqnarray}      
\label{30}
n^i=e^{-\Phi}(\eta^i+\Omega_{LNRF}\xi^i),\\
\Phi=\frac{1}{2}\ln{[-(\eta^i+\Omega_{LNRF}
\xi^i})(\eta_{i}+\Omega_{LNRF}
\xi_i)],
\end{eqnarray}
which corresponds to the \mbox{4-velocity} field of the locally non-rotating frames moving along circular orbits with the angular velocity $d\phi/dt=\Omega_{LNRF}$, defines the special observers \cite{Abr-Nur-Wex:1995:CLAQG:}. Since the vector field $n^i$ is hypersurface orthogonal, the unit vector $\tau^i$, used in the decomposition (\ref{8}), must be located in the hypersurface. 
On the other hand, the \mbox{4-velocity}     
\begin{eqnarray}
\label{31}
u^i=e^{-A}(\eta^i+\Omega\xi^i),\\
A=\frac{1}{2}\ln{[-(\eta^i+\Omega
\xi^i})(\eta_{i}+\Omega
\xi_i)],
\end{eqnarray}
corresponding to the general circular motion of a test particle with an angular velocity $d\phi/dt=\Omega\neq\Omega_{LNRF}$, is not hypersurface orthogonal. Moreover, the circular motion is directed along the Killing vector $\xi^i$, i.e.,
\begin{eqnarray}
\label{32}
\tau^i=(\xi^k\xi_k)^{-1/2}\xi^i.
\end{eqnarray}
Then, by using expressions (\ref{8}) and (\ref{31}), the Lorentz factor and velocity are given by the relations
\begin{eqnarray}
\label{33}
\gamma=e^{\Phi-A},\quad 
v=e^{-\Phi}(\xi^{i}\xi_{i})^{1/2}(\Omega-\Omega_{LNRF})
\end{eqnarray}
and components of the inertial forces (\ref{13})-(\ref{16}) are given by the relations 
\begin{eqnarray}
\label{35}
G_k=-m\,\nabla_k\Phi,\\
\label{36} 
Z_k=m(\gamma v)^2\,\frac{1}{2}(\xi^i\xi_i)^{-1}e^{-2\Phi}[e^{2\Phi}\nabla_{k}{(\xi^i\xi_i)}-{\xi^i\xi_i}{\nabla_k e^{2\Phi}}],\\
\label{37}
C_k=m\gamma^2v\,(\xi^i\xi_i)^{-3/2}e^{-\Phi}[\xi^i\xi_i\nabla_{k}{(\eta^i\xi_i)}-\eta^i\xi_i\nabla_k{(\xi^i\xi_i)}],\\
\label{38}
E_k=m\,e^{\Phi}\gamma^3 u^i\nabla_i(v) \tilde{\tau}_k.
\end{eqnarray}
The 'Newtonian' character of the inertial forces formalism appears after introducing the quantities  
\begin{eqnarray}
\label{41}
\tilde{R}=(\xi^{i}\xi_{i})^{1/2}e^{-\Phi},\quad \tilde{\Omega}\equiv\Omega-\Omega_{LNRF},\quad\tilde{v}=\gamma v,  
\end{eqnarray}
where $\tilde{\Omega}$ is angular velocity relative to the LNRF. Then we can write $v=\tilde{R}\tilde\Omega$ and the inertial forces (\ref{35})-(\ref{38}) can be rewritten to the form
\begin{eqnarray}
\label{35a}
G_k=-m\nabla_k\Phi,\\
\label{36a} 
Z_k=m\tilde{v}^2\,\tilde{R}^{-1}\nabla_{k}\tilde{R},\\
\label{37a}
C_k=-m(1+\tilde{v}^2)^{1/2}\tilde{v}\,\tilde{R}\nabla_k\Omega_{LNRF},\\
\label{38a}
E_k=m(1+\tilde{v}^2)^{3/2}\,e^{\Phi}\tilde{R}u^i\nabla_i(\Omega) \tilde{\tau}_k.
\end{eqnarray}
The quantity $\tilde{R}$ corresponds to the so-called radius of gyration, defined by the relation      
\begin{eqnarray}
\label{63h}
\tilde{R}=\sqrt{\frac{L}{\tilde{E}\tilde{\Omega}}},
\end{eqnarray}
generalizing in a special way its definition introduced and discussed for the static Schwarzschild spacetimes in \cite{Abr-Mil-Stu:1993:PHYSR4:}, \cite{Abr-Nur-Wex:1995:CLAQG:}.
The quantity $L=u^i\xi_i$ is the conserved specific angular momentum and $\tilde{E}=-u^i\iota_i$ is a scalar called 'pseudo-energy' of the particle. Since the vector field $\iota^i$ is not a Killing vector field, such 'pseudo-energy' is not conserved in general stationary and axially symmetric spacetimes. Of course, in the static spacetimes, $\tilde{E}$ is conserved, being defined in the standard way as $\tilde{E}=E=-u^i\eta_i$.    
The radius of gyration plays an important role in the theory of rotational effects in strong gravitational fields. The direction of increasing radius of gyration gives a preferred determination of the local outward direction relevant for the dynamical effects of rotation, whereas the direction becomes misaligned with the 'global' outward direction in strong fields \cite{Abr-Mil-Stu:1993:PHYSR4:}. 
%*************************************************************************************
\section{Kerr-de Sitter spacetimes}
In the standard Boyer-Lindquist coordinates $x^i=(t,r,\theta,\phi)$ and geometric units $\mbox{(c=G=1)}$, the line element of the KdS geometry is given by the relation
\begin{eqnarray}
\label{45}
ds^{2}=-\frac{\Delta_r}{I^2\rho^2}(dt-a\sin^2{\theta}d\phi)^2+\frac{\Delta_{\theta}\sin^2{\theta}}{I^2\rho^2}[adt-(r^2+a^2)d\phi]^2+\nonumber\\
\frac{\rho^2}{\Delta_r}dr^2+\frac{\rho^2}{\Delta_{\theta}}d\theta^2,
\end{eqnarray}
where
\begin{eqnarray}
\label{46}
\Delta_r=r^2-2Mr+a^2
    -\frac{1}{3}{\Lambda}r^2(r^2+a^2),\\
\label{47}
\Delta_{\theta}
  =1+\frac{1}{3}{\Lambda}a^2\cos^2{\theta},\\
\label{48}
I=1+\frac{1}{3}{\Lambda}a^2,\\
\label{49}
\rho^2=r^{2}+a^2\cos^{2}{\theta}. 
\end{eqnarray}
The mass $M$, specific angular momentum $a$, and cosmological constant $\Lambda$ are parameters of the spacetime. It is convenient to introduce a dimensionless cosmological parameter
\begin{eqnarray}
\label{50}
y=\frac{1}{3}\Lambda M^2.
\end{eqnarray}
For simplicity, we put $M=1$ hereafter. Equivalently, also the coordinates $t$, $r$, the line element $ds$, and the parameter $a$ are expressed in units of $M$ and become dimensionless.  

The KdS spacetimes, being stationary and axially symmetric, admit both the Killing vector fields $\eta^i$ and $\xi^i$, whereas  $\eta^i\eta_i=g_{tt}$,  
$\xi^i\xi_i=g_{\phi\phi}$ and $\eta^i\xi_i=g_{t\phi}$. The only intrinsic singularity of the KdS solution is the ring singularity in the equatorial plane and it is given by the relation $\rho=0$. Stationary regions of the spacetimes are determined by the relation $\Delta_r(r;a^2,y)\geq 0$ and limited by the inner and outer black-hole horizons at $r_{h-}$ and $r_{h+}$ and by the cosmological horizon at $r_{c}$, which are real roots of the equation $\Delta_r(r;a^2,y)=0$. Then spacetimes containing three horizons are black-hole spacetimes, while spacetimes containing one horizon (the cosmological horizon exists for any choice of the spacetime parameters) are naked-singularity spacetimes. Spacetimes with two horizons are extreme black-hole or extreme naked-singularity spacetimes \cite{Stu-Sla:2004:PHYSR4:}. 

In order to obtain the number of horizons in dependence on the rotational and cosmological parameters $a^2$ and $y$, we use the 'Chinese boxes' technique that will be used later in more complex discussions of the properties of the inertial forces and embedding diagrams.
In the KdS spacetimes, functions of the radius and two spacetime parameters are usually considered, and the routine is realized in two steps. A property of a function can be examined by investigation of properties of another function of the radius and of the lower number of parameters, i.e., of the only spacetime parameter. These properties of the function can be examined by investigation of properties of other functions, but now, functions of the radius only.  
Thus the properties of these functions can be determined by using a common way of investigation of functions dependent on one variable. Finally, the KdS spacetimes can be classified according to the initial property of the initial function. 

\begin{figure}
\begin{center}
\includegraphics[width=\hsize,clip=]{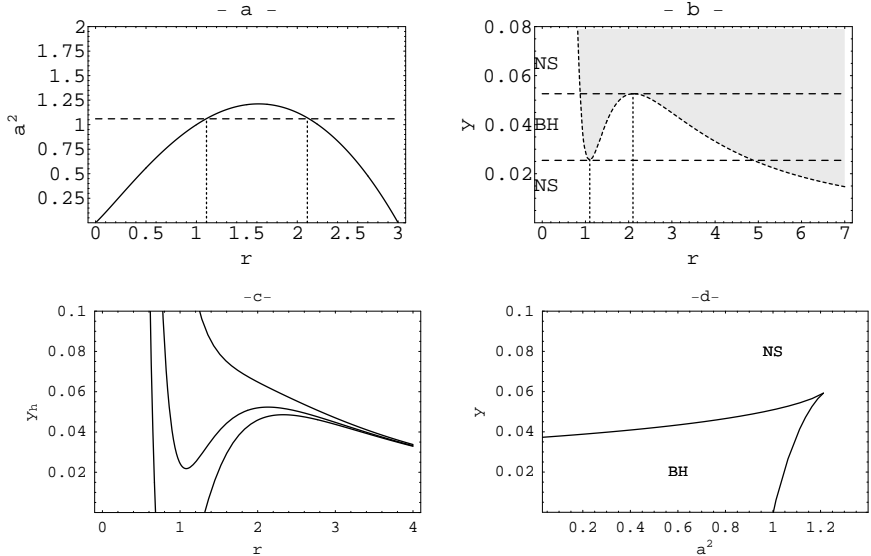}
\end{center}
\caption{(a) Characteristic function $a^2_{he}(r)$ governing the location of extrema of the function $y_h(r;a^2)$. For given $a^2$, the extrema of $y_h(r;a^2)$ are determined by solutions of $a^2=a^2_{he}(r)$ (note the dashed line $a^2=1.06$ and compare with figure~\ref{Fig:1}b).     
(b) Function $y_h(r;a^2)$ determining  the loci of event horizons of the KdS spacetimes and limiting the dynamic region where $y>y_h(r;a^2)$ (gray). The function is given for $a^2=1.06$. For given $a^2$ and $y>0$, the horizons are determined by solutions of $y=y_h(r;a^2)$. In the parameter line $(y)$ of the spacetimes, the extrema of $y_h(r;a^2)$  separate regions corresponding to the black-hole (BH) spacetimes and naked-singularity (NS) spacetimes for given values of the rotational \mbox{parameter $a^2$}. (c) Functions $y_h(r;a^2)$ in three specific cases; $a^2=0.9$ (with one positive and one negative extrema), $a^2=1.05$ (with two positive extrema) and $a^2=1.4$ (with no extrema). (d) Functions $y_{h,min}(a^2)$ and $y_{h,max}(a^2)$ separating the parametric plane $(a^2,y)$ into regions corresponding to the BH and NS spacetimes.}   
\label{Fig:1}
\end{figure}

In this case, it follows from the relation $\Delta_r(r;a^2,y)=0$ that for given $a^2$ and $y>0$ (repulsive cosmological constant), the loci of horizons are given by solutions of the equation  
\begin{eqnarray}
\label{54}
y=y_h(r;a^2)
\equiv\frac{r^2-2r+a^2}{r^2(r^2+a^2)}.
\end{eqnarray}
The reality condition $\Delta_r(r;a^2,y)>0$ can be then rewritten in the form  
\begin{eqnarray}
\label{55a}
0<y<y_h(r;a^2).\end{eqnarray}
The asymptotic behaviour of the function $y_h(r;a^2)$ is given by
$y_h(r\rightarrow\infty;a^2)\rightarrow+0$ and
$y_h(r\rightarrow 0;a^2)\rightarrow \infty$. The local extrema of  $y_h(r;a^2)$ are determined (due to the condition $\partial_r y_h(r;a^2)=0$) by the relation
\begin{eqnarray}
\label{58}
a^2(r)=a^2_{he}(r)\equiv\frac{1}{2}(-2r^2+\sqrt{8r+1}r+r),
\end{eqnarray}
the maximum of the function $a^2_{he}(r)$ is located at $r\doteq1.6160$ and takes the value $a^2_{he,max}\doteq1.2120$ (see figure~\ref{Fig:1}a).
We can distinguish three different types of behaviour of $y_h(r;a^2)$ (see figure~\ref{Fig:1}c).
\begin{itemize}
\item For $a^2<a^2_{he,max}$, $y_h(r;a^2)$ has two local extrema $y_{h,min}(a^2)$ and $y_{h,max}(a^2)$ determined by relations (\ref{54}) and (\ref{58}) 
(see figure~\ref{Fig:1}b). The black-hole spacetimes exist for $y_{h,min}(a^2)\leq y< y_{h,max}(a^2)$ and $y>0$, while the naked-singularity 
spacetimes exist for $0<y<y_{h,min}(a^2)$ or $y\geq y_{h,max}(a^2)$. 
\item For $a^2=a^2_{he,max}$, the extrema $y_{h,min}(a^2)$ and $y_{h,max}(a^2)$ 
coincide at $y_{h,crit}\doteq0.0592$, which is the limiting value for the black-hole spacetimes. 
\item For $a^2>a^2_{he,max}$, $y_h(r;a^2)$ has no 
extrema and there are only the naked-singularity spacetimes for any positive value of $y$. 
\end{itemize}

The parameter plane $(a^2,y)$ 
separated by the functions $y_{h,min}(a^2)$ and $y_{h,max}(a^2)$ into the regions corresponding to the black-hole and 
naked-singularity spacetimes is illustrated in figure \ref{Fig:1}d.  
%***************************************************************************
\section{Behaviour of inertial forces for circular motion} 
Assuming a test particle circular motion with the \mbox{4-velocity} (\ref{31}) at a fixed radius and latitude in the reference frame $x^i=(t,\phi,r,\theta)$ in the stationary and axially symmetric KdS  spacetimes, there are always zero time and azimuthal components of the gravitational, centrifugal and Coriolis forces (\ref{35})-(\ref{37}), and zero radial and latitudinal components of the Euler force (\ref{38}). 
The centrifugal, Coriolis and Euler forces vanish for $v=0$ and moreover, the Euler force also vanishes in the case of the uniform motion $v=const$.   
However, due to the behaviour of the velocity independent parts of the forces, they can vanish (and change their sign) at some
radii and latitudes independently of the velocity. The same can happen for the gravitational force which is velocity independent by definition. Note that we pay no more attention to the Euler force, because of its vanishing in the case of the uniform motion considered in the following.     
Then it is convenient to rewrite the inertial forces to the form
\begin{eqnarray}
G_k=m\,\mathcal{G}_k,\quad Z_k=m\tilde{v}^2\,\mathcal{Z}_k, \quad C_k=m\tilde{v}(1+\tilde{v}^2)^{1/2}\,\mathcal{C}_k,
\end{eqnarray}
whereas the non-vanishing velocity independent parts of the accelerations are given by the relations
\begin{eqnarray}
\label{61}
\mathcal{G}_r=\frac{r\Delta_r+\mathcal{A}\Delta_{\theta}(\mathcal{AB}+2r\Delta_r)\rho^2}{\mathcal{D}\Delta_r\rho^2},
\end{eqnarray}
\begin{eqnarray}
\label{62}
\mathcal{Z}_r=\frac{a^2\sin^2{\theta}\Delta_r(2r\Delta_r+\mathcal{B}\rho^2)+\mathcal{A}\Delta_{\theta}[2r(2\rho^2-\mathcal{A})\Delta_r+\mathcal{AB}\rho^2]}{\mathcal{D}\Delta_r\rho^2},
\end{eqnarray}
\begin{eqnarray}
\label{63}
\mathcal{C}_r=\frac{2a\sin{\theta}[r\mathcal{A}^2\Delta_{\theta}-\mathcal{AB}\rho^2-r\Delta_r(\mathcal{A}+\rho^2)]\sqrt{\Delta_r\Delta_{\theta}}}{\mathcal{D}\Delta_r\rho^2},
\end{eqnarray}
\begin{eqnarray}
\label{61t}
\mathcal{G}_{\theta}=\frac{a^2\sin{\theta}\cos{\theta}(\mathcal{D}\Delta_{\theta}-I\Delta_r\rho^2)}{\mathcal{D}\Delta_{\theta}\rho^2},
\end{eqnarray}
\begin{eqnarray}
\label{62t}
\mathcal{Z}_{\theta}=\frac{\cot{\theta}[2\mathcal{AD}\Delta_{\theta}-I(a^2\Delta_{\theta}+a^2\sin^2{\theta}\Delta_r)\rho^2]}{\mathcal{D}\Delta_{\theta}\rho^2},
\end{eqnarray}
\begin{eqnarray}
\label{63t}
\mathcal{C}_{\theta}=\frac{4a^3r\cos{\theta}\sin{\theta}\sqrt{\Delta_r\Delta_{\theta}}}{\mathcal{D}\Delta_{\theta}\rho^2},
\end{eqnarray}
where 
\begin{eqnarray}
\label{630}
\mathcal{A}=r^2+a^2,\quad\mathcal{B}=1-r+yr(a^2+2r^2),\\
\mathcal{D}=\mathcal{A}^2\Delta_{\theta}-a^2\Delta_r\sin^2{\theta}.
\end{eqnarray}
The presented components of the accelerations are considered to be functions of two variables $r$ and $\theta$, and two spacetime parameters $a^2$ and $y$.
There is $\mathcal{D}>0$ and $\Delta_{\theta}>0$ for all the values of $r$ and $\theta$. Thus the functions are well defined throughout the stationary regions.
The functions diverge at the radii of event horizons and in the ring singularity in the equatorial plane. 
Moreover, the latitudinal component of the velocity independent part of the centrifugal acceleration diverges at the axis of symmetry ($\theta=0$). 

In the case of the equatorial motion ($\theta=\pi/2$), latitudinal components of all the forces vanish. For given values of the rotational and cosmological parameters $a^2$ and $y>0$, the radii of circular orbits where the velocity independent parts of the radial components of the gravitational, centrifugal and Coriolis accelerations, i.e., $\mathcal{G}_r(r;a^2,y)$, $\mathcal{Z}_r(r;a^2,y)$ and $\mathcal{C}_r(r;a^2,y)$, vanish are given by zero points of the numerators of relations (\ref{61})-(\ref{63}), provided that $y<y_h(r;a^2)$ because of the reality condition.
It is immediately clear from relation (\ref{63}) that $\mathcal{C}_r(r;a^2,y)$ does not vanish at all. 
Moreover, since the zero points of the functions $\mathcal{G}_r(r;a^2,y)$ and $\mathcal{Z}_r(r;a^2,y)$ enable changes of their sign, the gravitational and centrifugal forces also can change their orientation on such orbits. 
Detailed discussion of vanishing of the functions $\mathcal{G}_r(r;a^2,y)$ and $\mathcal{Z}_r(r;a^2,y)$ is given in \cite{Kov-Stu:2004:RAGtime4and5:}, \cite{Kov-Stu:2006:IJMPA:}. Here we present only an updated summary of the results. However, the radii of circular orbits where $\mathcal{Z}_r(r;a^2,y)$ vanishes are closely related to embedding diagrams of the equatorial plane of the optical geometry and will be discussed in more details \mbox{in section \ref{Emb}}.  

\begin{figure}
\begin{center}
\includegraphics[width=0.7\hsize,clip=]{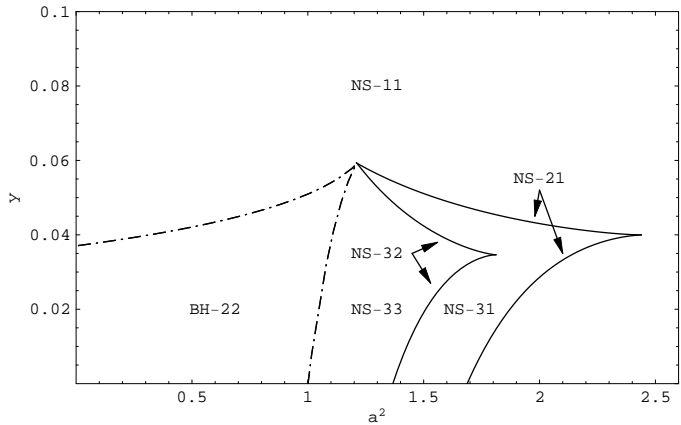}
\end{center}
\caption{Classification of the KdS spacetimes according to the number of circular orbits where $\mathcal{G}_r(r;a^2,y)=0$ (first digit) and $\mathcal{Z}_r(r;a^2,y)=0$ (second digit). In the parameter plane $(a^2,y)$, the functions $y_{h,max}(a^2)$ (upper dashed-dotted) and $y_{h,min}(a^2)$ (lower dashed-dotted) determining the extrema of the function $y_h(r;a^2)$ separate regions corresponding to the black-hole spacetimes BH-22 and to the naked-singularity spacetimes. The left solid and right solid curves determine the naked-singularity spacetimes corresponding to the classes \mbox{NS-32} and \mbox{NS-21} and separate the naked-singularity spacetimes into regions  corresponding to the classes \mbox{NS-11}, \mbox{NS-21}, \mbox{NS-31}, \mbox{NS-32} and \mbox{NS-33}.}   
\label{Fig:2}
\end{figure}

\begin{figure}
\includegraphics[width=0.9\hsize,clip=]{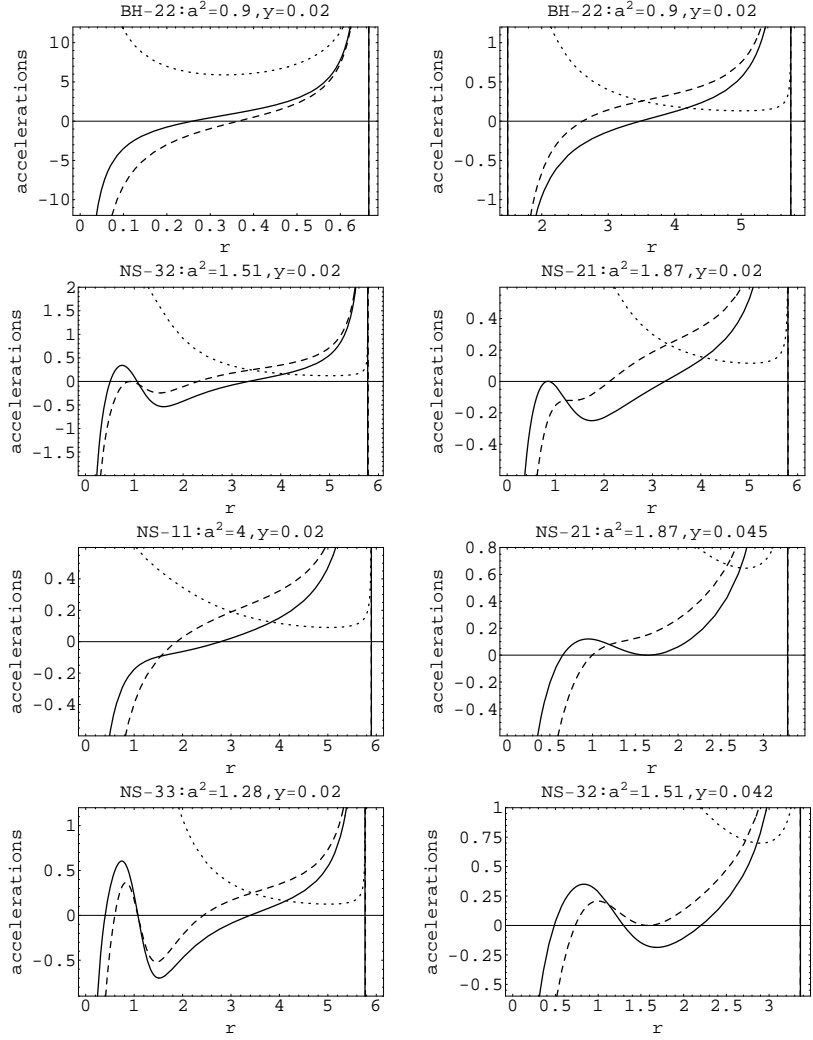}
\caption{Gravitational acceleration $\mathcal{G}_r$ (solid) and velocity independent parts of the  centrifugal $\mathcal{Z}_r$ (dashed) and Coriolis $\mathcal{C}_r$ (dotted) accelerations of particles moving along equatorial circular orbits in the KdS spacetimes. There are examples of qualitatively different types of behaviour of $\mathcal{G}_r(r;a^2,y)$ and $\mathcal{Z}_r(r;a^2,y)$ corresponding to the classification of the spacetimes according to the number of circular orbits where $\mathcal{G}_r(r;a^2,y)=0$ and $\mathcal{Z}_r(r;a^2,y)=0$. Positive (negative) parts of $\mathcal{G}_r(r;a^2,y)$ and $\mathcal{Z}_r(r;a^2,y)$  determine repulsive (attractive) forces oriented outwards (towards) the singularity at $r=0$. The positive function $\mathcal{C}_r(r;a^2,y)$ determines the repulsive (attractive) Coriolis force in the case of the corotating orbits with $\tilde{v}>0$ (counterrotating orbits with $\tilde{v}<0$). The zero points of the functions determine the radii of circular orbits where the forces vanish independently of the velocity. The vertical lines denote the radii of event horizons.}  
\label{Fig:3}
\end{figure}

There are one class of the KdS black-hole spacetimes and five classes of the naked-singularity spacetimes differing in the number of circular orbits where $\mathcal{G}_r(r;a^2,y)$ vanishes and orbits where $\mathcal{Z}_r(r;a^2,y)$ vanishes (see figure~\ref{Fig:2}).     
Examples of different types of behaviour of the functions are illustrated in figure~\ref{Fig:3}. 
%*********************************************************************** 
\section{Relativistic dynamics in terms of inertial forces}
In few examples we show how the results obtained by relatively complex standard general relativistic methods could be easily and straightforwardly obtained in the framework of the inertial forces approach. Therefore we believe that the method of inertial forces, combined with standard approaches, could bring in an effective way some new results in modelling accretion disks. 
\subsection{Equatorial circular motion}
Acceleration necessary to keep a particle in an uniform circular motion in the equatorial plane ($\theta=\pi/2$) can be expressed due to relation (\ref{12}) in the simple form \cite{Abr-Nur-Wex:1995:CLAQG:}, \cite{Stu-Hle-Jur:2000:CLAQG:} 
\begin{eqnarray}
\label{63a}
a^{\perp}_r(r,\tilde{v})=-\mathcal{G}_r-\tilde{v}^2\mathcal{Z}_r-\tilde{v}(1+\tilde{v}^2)^{1/2}\mathcal{C}_r,
\end{eqnarray}
where $-\infty<\tilde{v}<\infty$, while $-1<v<1$. This equation enables an effective discussion of the properties of both accelerated and geodesic motions. Remind that there are only radial components of the inertial forces non-vanishing and $a^{\perp}_{\theta}=a^{\perp}_{\phi}=0$ in the equatorial plane. 

For ultrarelativistic particles ($\tilde{v}^2 \gg 1$) we obtain from equation (\ref{63a}) the asymptotic equation  
\begin{eqnarray}
\label{63b}
a^{\perp}_r(r,\tilde{v})\approx-\mathcal{G}_r\mp\frac{1}{2}\mathcal{C}_r-\tilde{v}^2(\mathcal{Z}_r\pm\mathcal{C}_r),
\end{eqnarray}
for photons ($\tilde{v}^2\rightarrow\infty$) we obtain the circular geodesic motion determined by the equation
\begin{eqnarray}
\label{63c}
\mathcal{Z}_r\pm\mathcal{C}_r=0,
\end{eqnarray}
where the upper signs correspond to the corotating motion ($\tilde{v}>0$) and the lower signs to the counterrotating motion ($\tilde{v}<0$). Solving the equivalent equation $\mathcal{Z}_r^2-\mathcal{C}_r^2=0$, we directly obtain the relation for the photon circular orbits in the KdS spacetimes \cite{Stu-Sla:2004:PHYSR4:}
\begin{eqnarray}
\label{63cc}
y^2a^4r^3+2ya^2r^2(r+3)+r(r-3)^2-4a^2. 
\end{eqnarray} 
This equation will be discussed carefully in \mbox{section \ref{Emb}} in a relation with embedding diagrams of the optical geometry. The ultrarelativistic particles moving along the circular orbit at the radius of the photon circular orbits are kept by the acceleration 
\begin{eqnarray}
a^{\perp}_r(r)=-\mathcal{G}_r\mp\frac{1}{2}\mathcal{C}_r,
\end{eqnarray}
\label{63d}
which is velocity independent with an accuracy $O(\tilde{v}^{-2})$.

In static spacetimes, there is  $\mathcal{C}_r=0$, and at the radius of the photon circular orbit the acceleration of particles is   
\begin{eqnarray}
\label{63e}
a^{\perp}_r(r)=-\mathcal{G}_r,
\end{eqnarray}
and it is velocity independent exactly. Relation (\ref{63e}) is not limited to the case of ultrarelativistic particles, because $\mathcal{Z}_r=0$ at the radius of the photon circular orbit in static spacetimes.  
\subsection{Limitation of circular geodesics}
Velocities of particles moving along equatorial circular geodesics are due to relations (\ref{63a}) and $a^{\perp}_r(r)=0$ determined by the expression 
\begin{eqnarray}
\label{63f}
\tilde{v}_{\pm}=\pm[\frac{\frac{1}{2}\mathcal{C}_r^2-\mathcal{G}_r\mathcal{Z}_r\mp\frac{1}{2}\mathcal{C}_r(\mathcal{C}_r^2-4\mathcal{G}_r\mathcal{Z}_r+4\mathcal{G}_r^2)^{1/2}}{\mathcal{Z}_r^2-\mathcal{C}_r^2}]^{1/2}.
\end{eqnarray}
It is clear that for the geodesical circular motion at a fixed radius, the condition  
\begin{eqnarray}
\label{63g}
\mathcal{C}_r^2-4\mathcal{G}_r\mathcal{Z}_r+4\mathcal{G}_r^2>0
\end{eqnarray}
must be satisfied. 
In the equatorial plane of the KdS spacetimes, this condition directly implies the relation  
\begin{eqnarray}
\frac{4r^3y-4}{-r\Delta_r}\geq0,
\end{eqnarray}
giving the limitation for the circular geodesics in the form $r\leq y^{-1/3}\equiv r_{s}$, i.e., we immediately obtain the limit given by the so-called static radius \cite{Stu-Sla:2004:PHYSR4:}.

In static spacetimes, where $\mathcal{C}_r=0$, the reality condition for the geodesics is $\mathcal{G}_r\mathcal{Z}_r<0$, i.e., the gravitational and centrifugal forces must point in opposite directions. 
\subsection{Static equilibrium positions}
A test particle is thought to be in a static equilibrium position, if it is at the 'rest' relative to the reference coordinate system, i.e., $\phi=const$, $r=const$ and $\theta=const$, whereas its worldline is a geodesic, which means $a^{\perp}_r=a^{\perp}_{\theta}=a^{\perp}_{\phi}=0$. It is equivalent to the circular geodesical motion at a fixed radius and latitude, with $\Omega=0$, i.e.,
\begin{eqnarray}
\label{64}
v=v_s\equiv-\tilde{R}\Omega_{LNRF}.
\end{eqnarray} 
Defining $\tilde{v}_s=\gamma v_s$, we obtain two equilibrium conditions
\begin{eqnarray}
\label{65a}
\mathcal{G}_r+\tilde{v}_s^2\mathcal{Z}_r+\tilde{v_s}(1+\tilde{v_s}^2)^{1/2}\mathcal{C}_r=0,\\
\label{65b}
\mathcal{G}_{\theta}+\tilde{v}_s^2\mathcal{Z}_{\theta}+\tilde{v_s}(1+\tilde{v_s}^2)^{1/2}\mathcal{C}_{\theta}=0
\end{eqnarray}
that have to be satisfied simultaneously. 

In the equatorial plane ($\theta=\pi/2$), condition (\ref{65b}) is automatically satisfied because of the vanishing of latitudinal components of all the forces. By using relations \mbox{(\ref{61})-(\ref{63})}, condition (\ref{65a}) implies the equation 
\begin{eqnarray}
\label{66}
\frac{1-yr^3}{a^2-\Delta_r}=0,
\end{eqnarray}
i.e., the static equilibrium positions in the equatorial plane are really located at the static radius $r_s=y^{-1/3}$.
On the symmetry axis ($\theta=0$), condition (\ref{65b}) is also automatically satisfied and condition (\ref{65a}) implies the equation
\begin{eqnarray}
\label{67}
\frac{a^2-r^2+yr(a^2+r^2)^2}{\mathcal{A}\Delta_r}=0,
\end{eqnarray}
implicitly determining the static radii on the axis of symmetry derived in a standard way and discussed in \cite{Stu-Kov:2006:CLAQG:}.
\subsection{Perfect fluid toroidal configurations}
By projecting the conservation laws $\nabla_k T^{ik}=0$, where \mbox{ $T^{ik}=(p+\epsilon)u^i u^k+pg^{ik}$} is the stress-energy tensor for the perfect fluid, into the \mbox{3-dimensional} space of the special observers (\ref{3}) by using the projection tensor $h^i_j=\delta^i_j+n^i n_j$, and by using the condition
\begin{eqnarray}
u_i\nabla_k T^{ik}=0,
\end{eqnarray}
we arrive at the equation 
\begin{eqnarray}
\label{b4}
\gamma^2 v\tau_j (n^k+v\tau^k) \nabla_k p+\nabla_jp+n_j n^k\nabla_kp+(p+\epsilon){F'_j}^{\perp}=0.
\end{eqnarray}
Assuming now a circular motion of the baryotropic fluid, i.e., body with the equation of state $p=p(\epsilon)$ and $p\neq p(t,\phi)$, in the axially symmetric spacetimes, i.e., the vector $\tau^i$ and velocity $v$ given by relations (\ref{32}) and (\ref{33}), and $n^i$ given by relation (\ref{30}), we arrive at the relativistic Euler equation in terms of the inertial forces in the 'Newtonian' form
\begin{eqnarray}
\label{b5}
\nabla_jp+(p+\epsilon){F'_j}^{\perp}=0.
\end{eqnarray}
Of course, here, there are the only radial and latitudinal components of the inertial force \mbox{${F'_j}^{\perp}$ (\ref{12})} non-vanishing. Then, in the case of the uniform distribution of the specific angular momentum \mbox{$l=L/E=const$}, the integration of equation (\ref{b5}) %and relations (\ref{61})-(\ref{63t})
directly implies the equipressure surfaces, determined by the relation 
\begin{eqnarray}
\label{b6}
p(r,\theta)=\ln{[\frac{\rho^2\Delta_r\Delta_{\theta}\sin^2{\theta}}{I^2\Delta_{\theta}(r^2+a^2-al)^2\sin^2{\theta}-\Delta_r(l-a\sin^2{\theta})^2}]^{1/2}},
\end{eqnarray}
derived in a standard way in \cite{Sla-Stu:2005:CLAQG:}
%*******************************************************************8
\section{Embedding diagrams}
\label{Emb}
Properties of the optical reference geometry can be represented by the embedding of the equatorial plane into the \mbox{3-dimensional} Euclidean space with the line element expressed in the cylindrical coordinates $(\rho,z,\alpha)$ in the form  
\begin{eqnarray}
\label{1e}
d\sigma^2=d\rho^2+\rho^2d\alpha^2+dz^2.
\end{eqnarray}
It should be stressed that such embedding diagrams are very useful in a representation of some properties of the spacetime relevant in astrophysically important situations; especially behaviour of the centrifugal force, directly demonstrated by the embedding of the optical reference geometry, is crucial for understanding the equilibrium configurations of perfect fluid important in modelling accretion discs. Here, we shall follow the method developed in the case of Kerr-Newman geometry in \cite{Stu-Hle-Jur:2000:CLAQG:}. 
\subsection{Embedding formula}
The embedding diagram is characterized by the embedding formula $z=z(\rho)$ determining a surface in the Euclidean space with the line element 
\begin{eqnarray}
\label{2e}
dl_{(E)}^2=[1+{(\frac{dz}{d\rho})}^2]d\rho^2+\rho^2d\alpha^2,
\end{eqnarray}
isometric to the \mbox{2-dimensional} equatorial plane of the optical space determined by the line element 
\begin{eqnarray}
\label{3e}
d\tilde{l}^2=\tilde{h}_{rr}dr^2+\tilde{h}_{\phi\phi}d\phi^2, 
\end{eqnarray}
where the metric coefficients of the optical geometry in the KdS background are given by the relations
\begin{eqnarray}
\label{4e}
\tilde{h}_{rr}=\frac{r(1+a^2y)^2[r^3+a^4ry+a^2(2+r+r^3y)]}{\Delta_r^2},\\
\label{4f}
\tilde{h}_{\phi\phi}=\frac{[r^3+a^4ry+a^2(2+r+r^3y)]^2}{r^2\Delta_r}.
\end{eqnarray}
The azimuthal coordinates can be identified ($\phi=\alpha$) and we can put $\rho^2=\tilde{h}_{\phi\phi}$. Thus the differential form of the embedding formula is governed by the relation 
\begin{eqnarray}
\label{5e}
(\frac{dz}{d\rho})^2=\tilde{h}_{rr}(\frac{dr}{d\rho})^2-1
\end{eqnarray}
It is convenient to transfer it into the parametric form $z(\rho)=z(r(\rho))$ with $r$ being the parameter, i.e., 
\begin{eqnarray}
\label{6e}
\frac{dz}{dr}=\pm\sqrt{\tilde{h}_{rr}-(\frac{d\rho}{dr})^2},
\end{eqnarray}
whereas the sign in this formula is irrelevant, leading to isometric surfaces \cite{Hle:2001:RAGtime2and3:}.

Using the relation $\rho^2=\tilde{h}_{\phi\phi}$ and the metric coefficients (\ref{4e}) and (\ref{4f}), we obtain the differential form of the embedding formula 
\begin{eqnarray}
\label{7e}
\frac{dz}{dr}=\pm\sqrt{\frac{L}{-\Delta_r^3r^4}},
\end{eqnarray}
where
\begin{eqnarray}
\label{8e}
L=Ay^4+By^3+Cy^2+Dy+E
\end{eqnarray}
and
\begin{eqnarray}
A=a^6r^6(a^2+r^2)^3,\\
B=a^4r^5(a^2+r^2)^2[3r^3+a^2(3r+10)],
\end{eqnarray}
\begin{eqnarray}
C=-a^2r^3(a^2+r^2)[-3r^7-a^2(6r^2+16r+9)r^3-\\\nonumber
a^4r(3r^2+12r+37)+4a^6],\\
D=r^2\{r^{10}+a^2r^6(3r^2+2r-18)+a^4r^3[r(3r^2-4r-16)+36]+\\\nonumber
a^6r[r(r^2-14r-18)+60]-4a^8(2r+5)\},\\
E=4a^8-4a^6r[r(r-3)+6]-3a^4(r-2)r^2[r(4r-3)+6]-\\\nonumber
2a^2r^5[r(6r-17)+18]+(9-4r)r^8.
\end{eqnarray}
%-------------------------------------------------------
\subsection{Features of embedding diagrams}
\subsubsection{Limits of embeddability}
From the differential form of the embedding formula (\ref{7e}), it is clear that due to the condition $(dz/dr)^2\geq0$, the equatorial plane of the optical geometry is not entirely embeddable into the 3-dimensional Euclidean space. The embeddable regions are determined by the condition $L(r;a^2,y)\leq0$, whereas the equality in this relation determines the limits of embeddability. 
The equation $L(r;a^2,y)=0$ implicitly defines two functions $y_{L\pm}(r;a^2)$, whereas $y_{L-}(r;a^2)<y_{L+}(r;a^2)$ for all values of $a^2$ and $r$. Instead of giving long explicit expressions for the functions $y_{L\pm}(r;a^2)$, we only present their properties and examples of different types of their behaviour (see figures~\ref{Fig:4} and \ref{Fig:5}). For given values of $a^2$ and $y$, the limits of embeddability are determined by solutions of the equations
\begin{eqnarray}
\label{10e}
y=y_{L\pm}(r;a^2). 
\end{eqnarray}
Because of the reality condition $\Delta_r(r;a^2,y)>0$ and the repulsive cosmological constant, the solutions are restricted by the condition
\begin{eqnarray}
\label{11e}
0<y<y_h(r;a^2).
\end{eqnarray} 
Thus the embeddable regions are given by the condition $0<y_{L-}(r;a^2)\leq y$ and $y\leq y_{L+}(r;a^2)$. 

The zero points of the functions $y_{L\pm}(r;a^2)$ are determined by the condition $E(r;a^2)=0$ (see equation (\ref{8e})), i.e., by solutions the equation
\begin{eqnarray}
\label{12e}
4a^8-4a^6r[r(r-3)+6]-3a^4(r-2)r^2[r(4r-3)+6]-\nonumber\\
2a^2r^5[r(6r-17)+18]+(9-4r)r^8=0,
\end{eqnarray}    
implicitly defining functions $a^2_{L0\pm}(r)$. The function $a^2_{L0+}(r)$ has two local maxima $a^2_{L0+,max1}\doteq1.1354$ and $a^2_{L0+,max2}\doteq1.0754$ located at $r\doteq0.6895$ and at $r=1.3317$. There is also one local minimum $a^2_{L0+,min}=1$ located at $r=1$. On the other hand, the function $a^2_{L0-}(r)$ has no local extrema. 

The number of solutions of equation (\ref{10e}) depends on the number of extrema of the functions $y=y_{L\pm}(r;a^2)$. They are governed by the functions $a^2_{Le1}(r)$ and $a^2_{Le2\pm}(r)$, which are defined implicitly by eliminating $y$ from the equations $L(r;a^2,y)=0$ and $\partial_r L(r;a^2,y)=0$.  Since these expressions are too large to be presented here, we only show behaviour of the functions $a^2_{Le1}(r)$ and $a^2_{Le2\pm}(r)$ (see figure~\ref{Fig:4}). 
The function $a^2_{Le1}(r)$ is identical with the function $a^2_{he}(r)$ (see relation (\ref{58})) governing the extrema of the function $y_h(r;a^2)$. Thus two extrema of $y_{L\pm}(r;a^2)$ coalesce with two extrema of $y_h(r;a^2)$. These two extrema are the only common points of $y_{L\pm}(r;a^2)$ and $y_h(r;a^2)$, and there is $y_{L\pm}(r;a^2)<y_{h}(r;a^2)$ except these points. The maximum of the function $a^2_{Le1}(r)$ takes the value $a^2_{Le1,max}\doteq1.2120$ and is located at $r\doteq1.6160$. The common point of $a^2_{Le1}(r)$ and $a^2_{L0+}(r)$ coalesces with the local minimum of $a^2_{L0+}(r)$ and divides the function $a^2_{Le1}(r)$ into two parts governing positive (relevant for the classification) and negative (irrelevant) extrema of $y_{L\pm}(r;a^2)$.
The function $a^2_{Le2+}(r)$ has local maximum $a^2_{Le2+,max}\doteq1.4706$ located at $r\doteq1.0961$ and local minimum $a^2_{Le2+,min}\doteq1.0683$ located at $r\doteq1.3289$. Two of three common points of $a^2_{Le2+}(r)$ and $a^2_{L0+}(r)$ coalesce with local maxima of $a^2_{L0+}(r)$ and divide the  function $a^2_{Le2+}(r)$ into parts governing positive and negative extrema of $y_{L\pm}(r;a^2)$. The function $a^2_{Le2-}(r)$ completely governs negative local extrema of $y_{L\pm}(r;a^2)$, which we do not consider here.

All the characteristic functions $a^2_{L0\pm}(r)$, $a^2_{Le1\pm}(r)$ and $a^2_{Le2}(r)$ are illustrated in figure~\ref{Fig:4}. These functions enable us to understand the behaviour of the functions $y_{L\pm}(r;a^2)$ and classify the KdS spacetimes according to the number of embeddable regions.
\begin{figure}
\begin{center}\includegraphics[width=0.7\hsize]{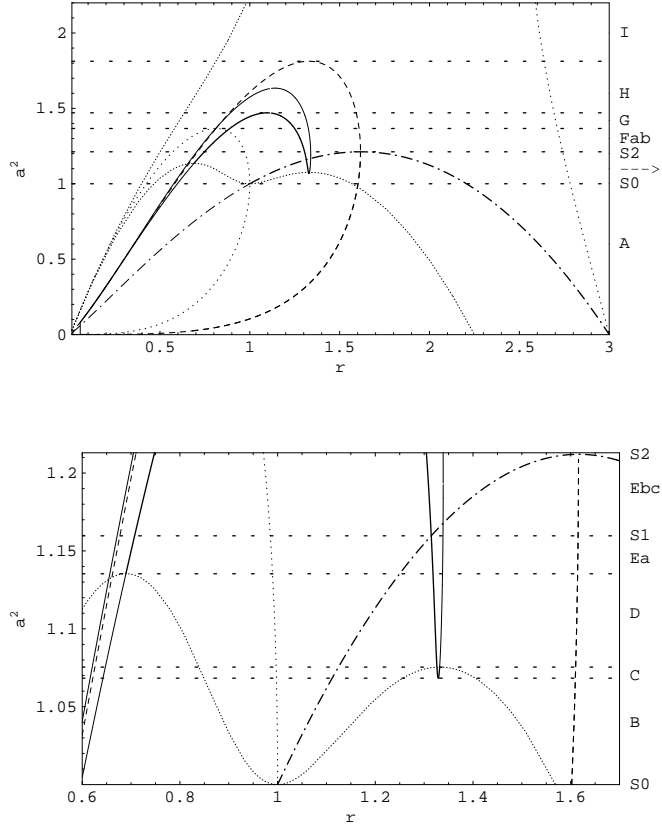}\end{center}
\caption{Characteristic functions $a^2_{L0\pm}(r)$ (closely dotted) governing zeros of the function $y_{L\pm}(r;a^2)$; $a^2_{Le1}(r)$ (dashed-dotted) and $a^2_{Le2\pm}(r)$ (solid) governing extrema of $y_{L\pm}(r;a^2)$; $a^2_{T0\pm}(r)$ (dotted) governing zeros of the function $y_T(r;a^2)$; $a^2_{Te\pm}(r)$ (dashed) governing extrema of $y_{T}(r;a^2)$. The thick parts of the dashed-dotted and solid curves govern positive extrema of $y_{L\pm}(r;a^2)$. In the parameter line $(a^2)$ of the spacetimes, the extrema and common points of the characteristic functions separate regions corresponding to different types of the behaviour of $y_{L\pm}(r;a^2)$ (see figure~\ref{Fig:5}).}
\label{Fig:4}  
\end{figure}
%-----------------------------------------------------------
\subsubsection{Turning points of embedding diagrams}
In general, the plasticity of the embedding diagrams is characterized by the different steepness and by the turning points of the diagrams. 
Clearly, the turning points of the embedding diagrams, i.e., their bellies and throats, are determined by the relation \mbox{$d\rho/dz=0$.} Because of the relation
\begin{eqnarray}
\label{13e}
\frac{dz}{d\rho}=\frac{dz}{dr}\frac{dr}{d\rho},
\end{eqnarray}
we have the condition $d\rho/dr=0$. 
Due to relations (\ref{7}), (\ref{41}) and $\rho^2=\tilde{h}_{\phi\phi}$, we can write  
$d\rho/dr=d\tilde{R}/dr=0$. 
Considering relation (\ref{36a}), it is clear that the turning points of the embedding diagrams are located at the radii where the velocity independent part of the centrifugal acceleration vanishes.

Since there is 
\begin{eqnarray}
\label{14e}
\frac{d\rho}{dr}=\{r^2\Delta_r^{3/2}\}^{-1}\{r^3a^4(a^2+r^2)y^2+r^2a^2[(2r+5)a^2+r^2(2r+3)]y+\nonumber\\
r^4(r-3)+ra^2[r(r-3)+6]-2a^4\},
\end{eqnarray}  
then for given values $a^2$ and $y$, the radii of turning points of the embedding diagrams are given by solutions of the equation
\begin{eqnarray}
\label{15e}
y=y_{T}(r;a^2)\equiv\{2a^2r^2(a^2+r^2)\}^{-1}\{-(2r+5)ra^2-r^3(2r+3)+\nonumber\\
\sqrt{r(a^2+3r^2)[8a^4+ra^2(16r+1)+r^3(8r+3)]}\}
\end{eqnarray}
restricted by the condition
\begin{eqnarray}
\label{16e}
0<y_{L-}(r;a^2)<y<y_{L+}(r;a^2).
\end{eqnarray} 

The zero points of the function $y_T(r;a^2)$ are determined by solutions of the equation 
\begin{eqnarray}
\label{16f}
r^4(r-3)+ra^2[r(r-3)+6]-2a^4=0,
\end{eqnarray}
which we consider as an implicit form of the functions $a^2_{T0\pm}(r)$. 
The maximum of this function takes the value $a^2_{T0\pm,max}\doteq 1.3667$ located at $r\doteq0.8116$. 

The number of solutions of equation (\ref{15e}) depends on the number of the extrema of the function $y_T(r;a^2)$. They are determined by solutions of the equation
\begin{eqnarray}
\label{80z}
\{\sqrt{a^4r^3(a^2+3r^2)[8a^4+r(16r+1)a^2+r^3(8r+3)]}\}^{-1}\times\nonumber\\
\{r[12a^8+r(48r+1)a^6+3r^3(24r+1)a^4+3r^5(16r+1)a^2+\nonumber\\
3r^7(4r+3)]a^2\}-5a^4-12r^2a^2-3r^4=0,
\end{eqnarray}
which we consider as an implicit form of the functions $a^2_{Te\pm}(r)$ with the maximum $a^2_{Te+,max}\doteq 1.8126$ located at $r\doteq1.3285$.

The characteristic functions $a^2_{Te\pm}(r)$ and $a^2_{T0\pm}(r)$ are illustrated in figure~\ref{Fig:4} and enable us to understand the behaviour of the functions $y_{T}(r;a^2)$ and classify the KdS spacetimes according to the number of turning points of the embedding diagrams.
%----------------------------------------     
\subsubsection{Classification}
\label{section}
The number of embeddable regions and turning points of the embedding diagrams is determined by the number of solutions of equations (\ref{10e}) and (\ref{15e}), which depends on the number of  extrema of the functions $y_{L\pm}(r;a^2)$ and $y_T(r;a^2)$.
Therefore we denote $y_{L-,max}(a^2)$ as the only maximum of $y_{L-}(r;a^2)$ which becomes positive (relevant for the classification) for some values of $a^2$ and $r$, $y_{L+,e1}(a^2)$, $y_{L+,e2}(a^2)$ as the extrema of $y_{L+}(r;a^2)$ coalescent with the minimum and maximum of the function $y_h(r;a^2)$; $y_{L+,e3}(a^2)$, $y_{L+,e4}(a^2)$ as the remaining two extrema of $y_{L+}(r;a^2)$. Finally we denote $y_{T,min}(a^2)$ and $y_{T,max}(a^2)$ as the minimum and maximum of the function $y_T(r;a^2)$.  

Using the characteristic functions $a^2_{Te\pm}(r)$, $a^2_{T0\pm}(r)$, $a^2_{L0\pm}(r)$, $a^2_{Le1}(r)$ and $a^2_{Le2\pm}(r)$, we can distinguish different types of behaviour of the functions $y_{L\pm}(r;a^2)$ and $y_{T}(r;a^2)$ differing in the number of their extrema satisfying the condition $0<y\leq y_h(r;a^2)$ (see figure~\ref{Fig:5}). The different types of the KdS spacetimes obtained this way are shown in table~\ref{Tab:1}.  
\begin{figure}
\begin{center}\includegraphics[width=1\hsize]{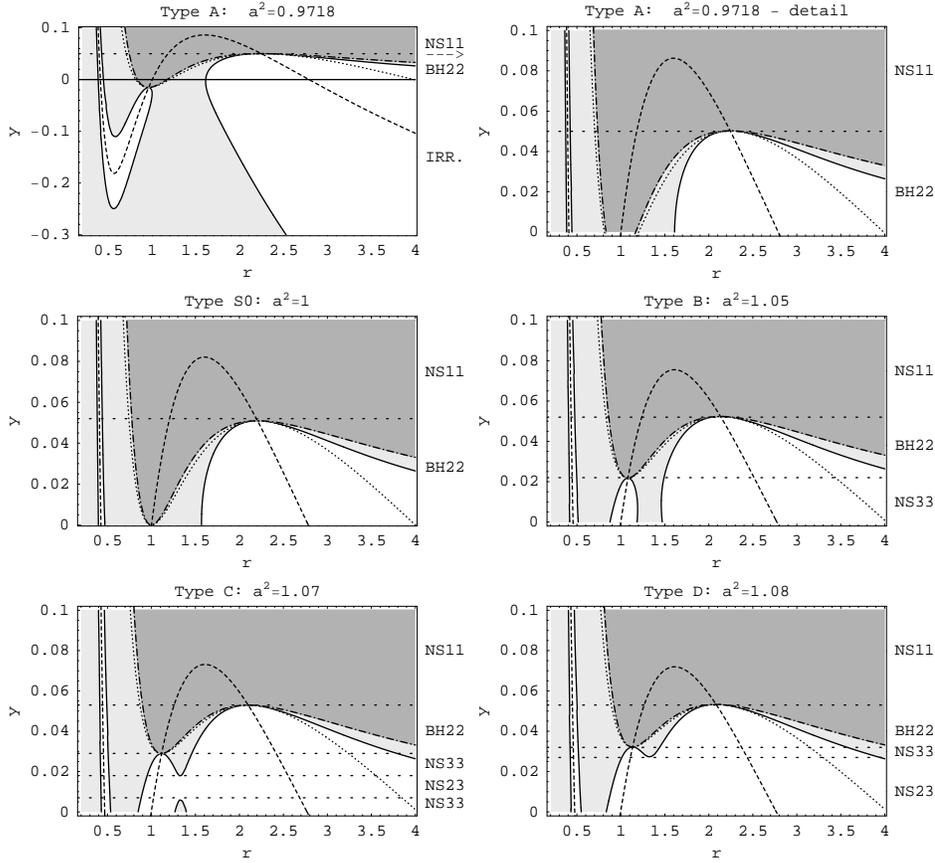}\end{center}
\caption{Functions $y_{L\pm}(r;a^2)$ (solid) limiting embeddable regions (white); $y_T(r;a^2)$ (dashed) determining turning points of embedding diagrams; $y_h(r;a^2)$ (dashed-dotted) determining  locations of event horizons and limiting the dynamic regions (gray). For illustration, the behaviour of the function  $y_{ph}(r;a^2)$ (dotted) governing radii of photon circular orbits is also given. For given $a^2$ and $y>0$, limits of embeddability are determined by solutions of $y=y_{L\pm}(r;a^2)$ that satisfy the condition $y<y_h(r;a^2)$. Turning points of the embedding diagrams are determined by solutions of $y=y_T(r;a^2)$ and embeddable photon circular orbits by solutions of $y=y_{ph}(r;a^2)$, whereas the solutions must satisfy the embeddability condition $y_{L-}(r;a^2)<y<y_{L+}(r;a^2)$. In the parameter line ($y$) of the KdS spacetimes, extrema of $y_{L\pm}(r;a^2)$ and $y_T(r;a^2)$ separate regions corresponding to different classes of the spacetimes differing in the number of embeddable regions (first digit) and turning points of the embedding diagrams (second digit). Note that in the case of the classes NS12 and NS22, the second digits exceptionally denote one turning point and one inflexion point of the diagrams.}
\label{Fig:5}
\addtocounter{figure}{-1}  
\end{figure}
\begin{figure}
\begin{center}\includegraphics[width=1\hsize]{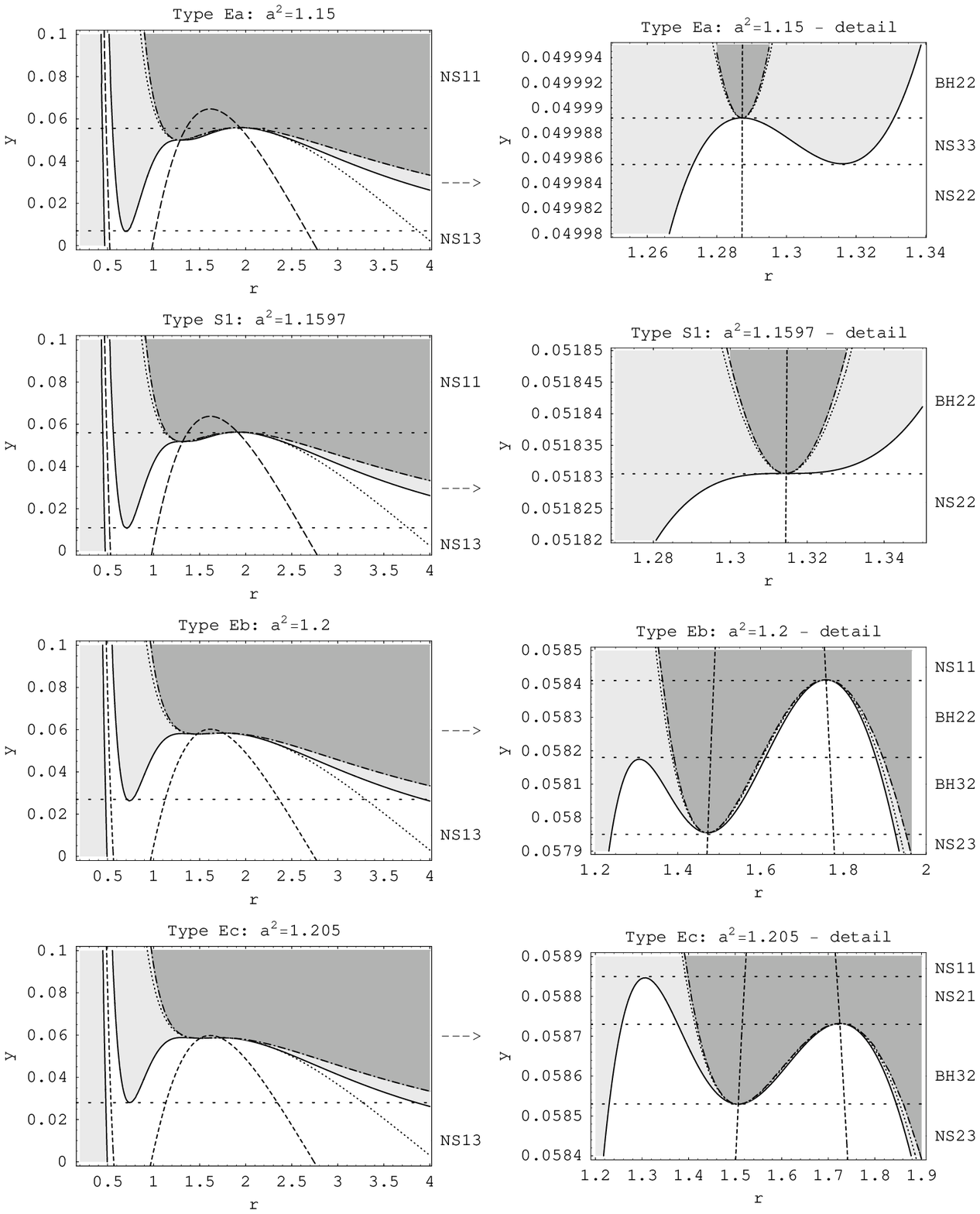}\end{center}
\caption{(Continued.)}
\addtocounter{figure}{-1} 
\end{figure}
\begin{figure}
\begin{center}\includegraphics[width=1\hsize]{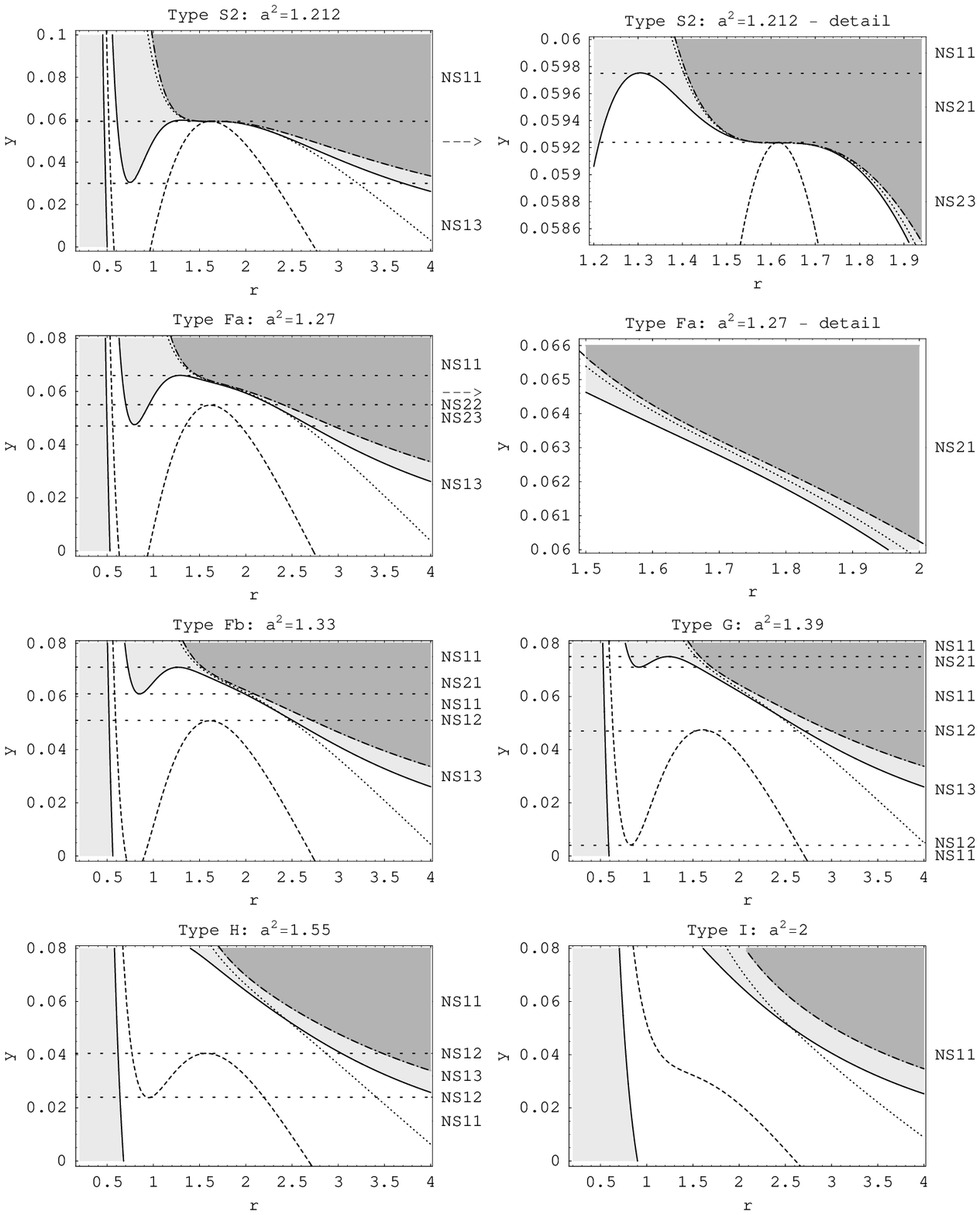}\end{center}
\caption{(Continued.)}
\end{figure}
\begin{table}
\caption{\label{Tab:1} Different types of the KdS spacetimes determined by corresponding different types  of behaviour of the functions $y_{L\pm}(r;a^2)$ and $y_{T}(r;a^2)$. The number of extrema $y_{L+,e1}(a^2)$, $y_{L+,e2}(a^2)$, $y_{L+,e3}(a^2)$, $y_{L+,e4}(a^2)$, $y_{L-,max}(a^2)$, $y_{T,min}(a^2)$ and $y_{T,max}(a^2)$ of the functions $y_{L\pm}(r;a^2)$ and $y_{T}(r;a^2)$ satisfying the condition $0<y\leq y_h(r;a^2)$ are subsequently expressed by the digits in the last column. The types Ea, Eb and Ec differ in the relation among the values of extrema $y_{L+,e2}(a^2)$, $y_{L+,e3}(a^2)$ and $y_{L+,e4}(a^2)$ and the types Fa and Fb differ in the relation among the values of extrema $y_{L+,e3}(a^2)$ and $y_{T,max}(a^2)$.}
\begin{minipage}{.4\linewidth}
\begin{indented}
\item\begin{tabular}{@{}lll}
\br
\bf{T}&\bf{Range of} $\bf a^2$&\bf{Extrema}\\ 
\hline
A&$(0,1\rangle$&0,1,0,0,0,0,0\\
S0&$1$&0,1,0,0,0,0,0\\
B&$(1,1.0683)$&1,1,0,0,0,0,0\\ 
C&$\langle 1.0683,1.0754)$&1,1,0,1,1,0,0\\ 
D&$\langle 1.0754,1.1354\rangle$&1,1,0,1,0,0,0\\
Ea&$(1.1354,1.1597)$&1,1,1,1,0,0,0\\
S1&$1.1597$&0,1,1,0,0,0,0\\
Eb&$(1.1597,1.2032)$&1,1,1,1,0,0,0\\
\br
\end{tabular}
\end{indented}
\end{minipage}
\begin{minipage}{.4\linewidth}
\begin{indented}
\item\begin{tabular}{@{}lll}
\br
\bf{T}&\bf{Range of} $\bf a^2$&\bf{Extrema}\\ 
\hline
Ec&$\langle1.2032,1.2120)$&1,1,1,1,0,0,0\\
S2&$1.2120$&0,0,1,1,0,0,1\\ 
Fa&$(1.2120,1.2938)$&0,0,1,1,0,0,1\\ 
Fb&$\langle1.2938,1.3667\rangle$&0,0,1,1,0,0,1\\ 
G&$(1.3667,1.4706)$&0,0,1,1,0,1,1\\
H&$\langle 1.4706,1.8126)$&0,0,0,0,0,1,1\\
I&$\langle 1.8126,\infty)$&0,0,0,0,0,0,0\\ 
\br
\\
\end{tabular}
\end{indented}
\end{minipage}
\end{table}
The classification of the KdS spacetimes according to the number of embeddable regions and turning points of the embedding diagrams can be now given in the following way. We step by step discuss the number of embeddable regions using the number of solutions of equation (\ref{10e}) restricted by condition (\ref{11e}) for given values of $a^2$ and $y$, and the number of turning points of embedding diagrams given by the solutions of equation (\ref{15e}), restricted by condition (\ref{16e}), for each type of the KdS spacetimes separately.

In the spacetimes of type 'A', for $0<y<y_{L+,e2}(a^2)$, there are four solutions of equation (\ref{10e}) satisfying condition (\ref{11e}) and then two embeddable regions. There are two solutions of  equation (\ref{15e}) satisfying condition (\ref{16e}), i.e., two turning points of the embedding diagrams. Moreover, there are three solutions of equation (\ref{54}) determining three even horizons and thus two stationary regions of the spacetimes. We denote the related class as BH22, alias the KdS black-hole spacetimes with two embeddable regions and two turning points. For $y \geq y_{L+,e2}(a^2)$, there are two solutions of equation (\ref{10e}) satisfying condition (\ref{11e}) and thus only one embeddable region. The only solution of equation (\ref{15e}) satisfying condition (\ref{16e}) determines one turning point of the embedding diagrams. Moreover, there is one solution of equation (\ref{54}), i.e., one cosmological horizon separating stationary region and dynamic region. We denote the related class as NS11, alias the KdS naked-singularity spacetimes with one embeddable region and one turning point. Note that although there are two horizons and two stationary regions in the case of $y=y_{L+,e2}(a^2)$, the outer stationary region coincides with the outer horizon where the geometry is not embeddable and thus we cannot consider it. Using the same way, we can sort the remaining types of the spacetimes obtaining the classification for all the KdS spacetimes (see table~\ref{Tab:2}).
\begin{table}
\caption{\label{Tab:2} Classification of the KdS spacetimes according to the number of embeddable regions (first digit) and turning points of the embedding diagrams (second digit).  
Limits of the ranges of the parameter $y$ (being functions of the parameter $a^2$) are illustrated in figure~\ref{Fig:8}.}
\begin{minipage}{.4\linewidth}
\begin{indented}
\item\begin{tabular}{@{}ll}
\br
\bf{Class}&\bf{Range of} $\bf y$\\
\hline
\it{Type A,}&\hspace{-3ex}\it{S0}\\
\hline
BH22&$(0;y_{L+,e2}(a^2))$\\
NS11&$\langle y_{L+,e2}(a^2);\infty)$\\
\hline
\it {Type B}&\\
\hline
NS33&$(0;y_{L+,e1}(a^2))$\\
BH22&$\langle y_{L+,e1}(a^2);y_{L+,e2}(a^2))$\\
NS11&$\langle y_{L+,e2}(a^2);\infty)$\\
\hline
\it{Type C}&\\
\hline
NS33&$(0;y_{L-,max}(a^2))$\\
NS23&$\langle y_{L-,max}(a^2);y_{L+,e4}(a^2))$\\
NS33&$(y_{L+,e4}(a^2);y_{L+,e1}(a^2))$\\
BH22&$\langle y_{L+,e1}(a^2);y_{L+,e2}(a^2)$\\
NS11&$\langle y_{L+,e2}(a^2);\infty)$\\
\hline
\it{Type D}&\\
\hline
NS23&$(0;y_{L+,e4}(a^2)\rangle$\\
NS33&$(y_{L+,e4}(a^2);y_{L+,e1}(a^2))$\\
BH22&$\langle y_{L+,e1}(a^2);y_{L+,e2}(a^2))$\\
NS11&$\langle y_{L+,e2}(a^2);\infty)$\\
\hline
\it{Type Ea}&\\
\hline
NS13&$(0;y_{L+,e3}(a^2)\rangle$\\
NS22&$(y_{L+,e3}(a^2);y_{L+,e4}(a^2)\rangle$\\
NS33&$(y_{L+,e4}(a^2);y_{L+,e1}(a^2))$\\
BH22&$\langle y_{L+,e1}(a^2);y_{L+,e2}(a^2))$\\
NS11&$\langle y_{L+,e2}(a^2);\infty)$\\
\hline
\it{Type S1}&\\
\hline
NS13&$(0;y_{L+,e3}(a^2)\rangle$\\
NS22&$(y_{L+,e3}(a^2);y_{h,min}(a^2))$\\
BH22&$\langle y_{h,min}(a^2);y_{L+,e2}(a^2))$\\
NS11&$\langle y_{L+,e2}(a^2);\infty)$\\
\hline
\it{Type Eb}&\\
\hline
NS13&$(0;y_{L+,e3}(a^2)\rangle$\\
NS23&$(y_{L+,e3}(a^2);y_{L+,e1}(a^2))$\\
BH32&$\langle y_{L+,e1}(a^2);y_{L+,e4}(a^2)\rangle$\\
BH22&$(y_{L+,e4}(a^2);y_{L+,e2}(a^2))$\\
NS11&$\langle y_{L+,e2}(a^2);\infty)$\\
\hline
\it{Type Ec}&\\
\hline
NS13&$(0;y_{L+,e3}(a^2)\rangle$\\
NS23&$(y_{L+,e3}(a^2);y_{L+,e1}(a^2))$\\
\br
\end{tabular}
\end{indented}
\end{minipage}
\hfill
\begin{minipage}{.58\linewidth}
\begin{indented}
\item\begin{tabular}{@{}ll}
\br
\bf{Class}&\bf{Range of} $\bf y$\\
\hline
\it{Type Ec}&\hspace{-3ex}-continued\\
\hline
BH32&$\langle y_{L+,e1}(a^2);y_{L+,e2}(a^2))$\\
NS21&$\langle y_{L+,e2}(a^2);y_{L+,e4}(a^2)\rangle$\\
NS11&$(y_{L+,e4}(a^2);\infty)$\\
\hline
\it{Type S2}&\\
\hline
NS13&$(0;y_{L+,e3}(a^2)\rangle$\\
NS23&$(y_{L+,e3}(a^2);y_{T,max}(a^2))$\\
NS21&$\langle y_{T,max}(a^2);y_{L+,e4}(a^2)\rangle$\\
NS11&$(y_{L+,e4}(a^2);\infty)$\\
\hline
\it{Type Fa}&\\
\hline
NS13&$(0;y_{L+,e3}(a^2)\rangle$\\
NS23&$(y_{L+,e3}(a^2);y_{T,max}(a^2))$\\
NS22&$y=y_{T,max}(a^2)$\\
NS21&$(y_{T,max}(a^2);y_{L+,e4}(a^2)\rangle$\\
NS11&$(y_{L+,e4}(a^2);\infty)$\\
\hline
\it{Type Fb}&\\
\hline
NS13&$(0;y_{T,max}(a^2))$\\
NS12&$y=y_{T,max}(a^2)$\\
NS11&$(y_{T,max}(a^2);y_{L+,e3}(a^2)\rangle$\\
NS21&$(y_{L+,e3}(a^2);y_{L+,e4}(a^2)\rangle$\\
NS11&$(y_{L+,e4}(a^2);\infty)$\\
\hline
\it{Type G}&\\
\hline
NS11&$(0;y_{T,min}(a^2))$\\
NS12&$y=y_{T,min}(a^2)$\\
NS13&$(y_{T,min}(a^2);y_{T,max}(a^2))$\\
NS12&$y=y_{T,max}(a^2)$\\
NS11&$(y_{T,max}(a^2);y_{L+,e3}(a^2)\rangle$\\
NS21&$(y_{L+,e3}(a^2);y_{L+,e4}(a^2)\rangle$\\
NS11&$(y_{L+,e4}(a^2);\infty)$\\
\hline
\it{Type H}&\\
\hline
NS11&$(0;y_{T,min}(a^2))$\\
NS12&$y=y_{T,min}(a^2)$\\
NS13&$(y_{T,min}(a^2);y_{T,max}(a^2))$\\
NS12&$y=y_{T,max}(a^2)$\\
NS11&$(y_{T,max}(a^2);\infty)$\\
\hline
\it{Type I}&\\
\hline
NS11&$(0;\infty)$\\
\br\\
\end{tabular}
\end{indented}
\end{minipage}
\end{table}

\begin{figure}
\begin{center}
\includegraphics[width=0.88\hsize]{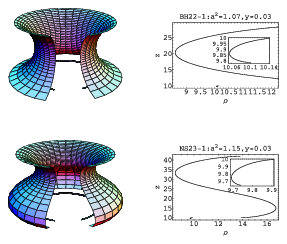}
\end{center}
\caption{Embedding diagrams of the optical reference geometry of the KdS spacetime of the classes BH$22-1$ and NS$23-1$.}
\label{Fig:9}
\end{figure}

\begin{figure}
\includegraphics[width=0.88\hsize]{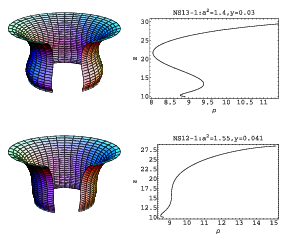}
\caption{Embedding diagrams of the optical reference geometry of the KdS spacetime of the classes  NS$13-1$ and NS$12-1$.}
\label{Fig:10}
\end{figure}

Now it is clear that the functions $y_{L+,e1}(a^2)$, $y_{L+,e2}(a^2)$, $y_{L+,e3}(a^2)$, $y_{L+,e4}(a^2)$, $y_{L-,max}(a^2)$, $y_{T,min}(a^2)$ and $y_{T,max}(a^2)$ separate the parametric plane $(a^2,y)$ into regions corresponding to different classes of the KdS black-hole spacetimes and naked-singularity  spacetimes differing in the number of embeddable regions and number of turning points of the embedding diagrams (see figure~\ref{Fig:8}).
Qualitatively different types of the embedding diagrams corresponding to the presented classification are illustrated in figures~\ref{Fig:9}-\ref{Fig:10}. 
%-------------------------------------------------------------------------
\subsection{Embeddability of photon circular orbits}
The radii of photon circular orbits are given by relation (\ref{63c}). Thus, of course, in the stationary (non-static) spacetimes, where $\mathcal{C}_r\neq0$,
the radii of photon circular orbits do not coalesce with the radii of the turning points of  embedding diagrams (radii of circular orbits where $\mathcal{Z}_r=0$), which holds for the static spacetimes, where $\mathcal{C}_r=0.$ 

\begin{figure}
\begin{center}\includegraphics[width=0.7\hsize]{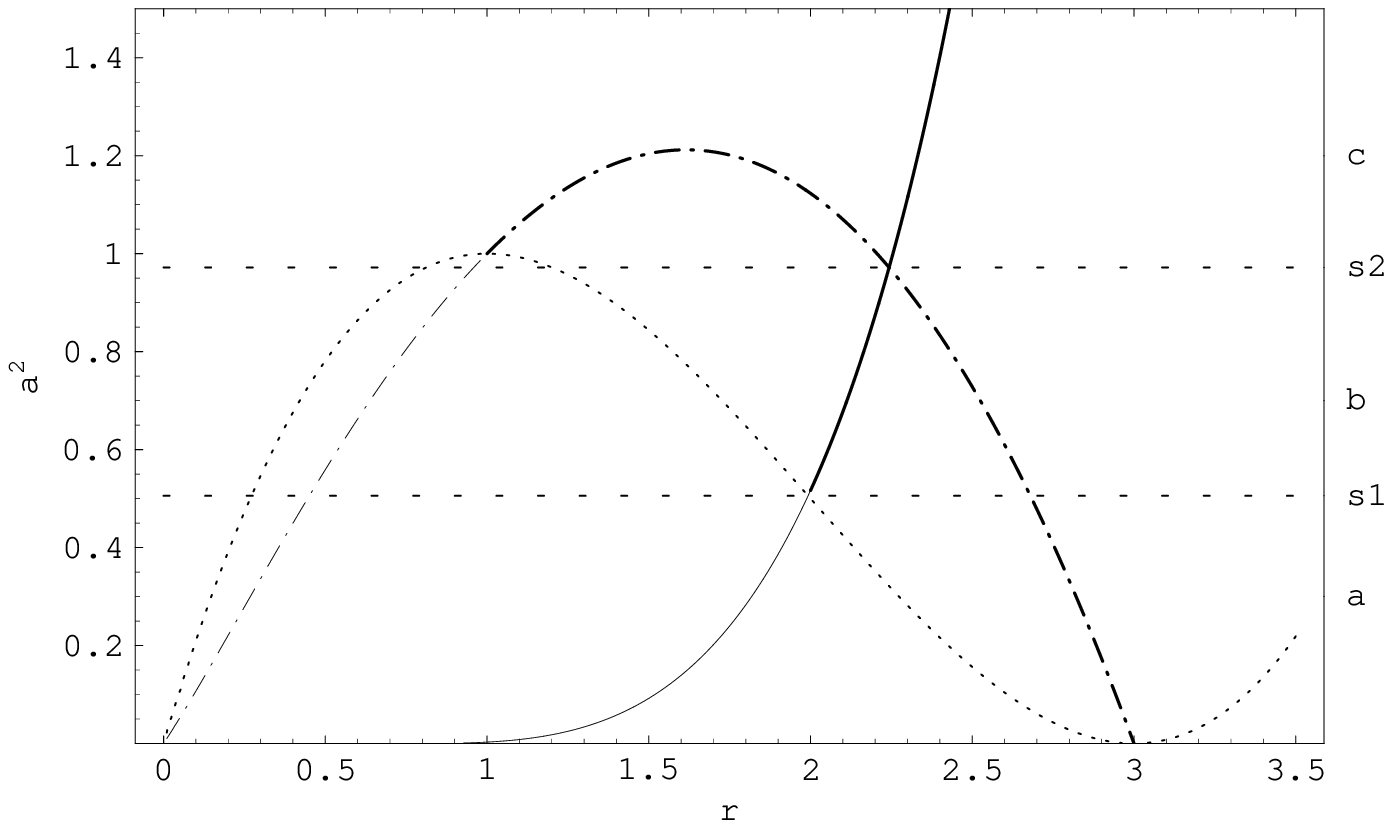}\end{center}
\caption{Characteristic functions $a^2_{phL}(r)$ (solid) governing common points of the functions $y_{ph}(r;a^2)$ and $y_{L\pm}(r;a^2)$; $a^2_{he}(r)$ (dashed-dotted) governing the extrema of $y_{ph}(r;a^2)$ coalescent with the extrema of the function $y_{h}(r;a^2)$ and with the other common points of $y_{ph}(r;a^2)$ and $y_{L\pm}(r;a^2)$; $a^2_{ph0}(r)$ (dotted) governing zeros of $y_{ph}(r;a^2)$. The physical relevant solutions are then denoted by the thick parts of the relevant curves. In the parameter line $(a^2)$ of the spacetimes, the common points of the characteristic functions separate regions corresponding to the different types of behaviour of $y_{ph}(r;a^2)$ (see figure~\ref{Fig:7}). }
\label{Fig:6}  
\end{figure}

For given values of $a^2$ and $y$, radii of circular photon orbits are determined (due to equation (\ref{63cc})) by solutions of the equation
\begin{eqnarray}
y=y_{ph}(r;a^2)\equiv\frac{-r^{1/2}(r+3)+2(3r^2+a^2)^{1/2}}{a^2r^{3/2}}
\label{ph1}
\end{eqnarray}
restricted by condition (\ref{55a}). The zero points of the function are determined by solutions of 
the equation
\begin{eqnarray}
a^2=a^2_{ph0}(r)\equiv\frac{1}{4}(r-3)^2r.
\label{ph10}
\end{eqnarray}
The number of circular orbits is determined by the extrema of the function, which coalesce with the extrema of the function $y_{h}(r;a^2)$. Thus they are governed by the function $a^2_{he}(r)$ given by equation (\ref{58}). 
Solutions of equation (\ref{ph1}) satisfying condition (\ref{16e}) then correspond to the embeddable orbits, number of which depends on the number of common points of $y_{ph}(r;a^2)$ and $y_{L\pm}(r;a^2)$. The points are determined by solutions of the equation
\begin{eqnarray}
\label{ph2}
[a^4+(r-3)r^3+a^2r(2r-1)]\{16a^{10}+16a^8(23-4r)r^2-\nonumber\\
9r^8(4r-3)^3+4a^4r^5[9+32(45-11r)r]-8a^6r^3(8r(8r-37)-9]+\nonumber\\
3a^2r^6[27-16r[9+r(32r-105)]]\}=0,
\end{eqnarray}
implicitly defining the function $a^2_{phL}(r)$, which determines a first kind of common points, and the function $a^2_{he}(r)$, determining the other kind, coalescing with the extrema of $y_{h}(r;a^2)$. 
All the characteristic functions $a^2_{phL}(r)$, $a^2_{he}(r)$ and $a^2_{ph0}(r)$ are illustrated in figure~\ref{Fig:6}. 
\subsubsection{Classification}
\begin{figure}
\begin{center}\includegraphics[width=1\hsize]{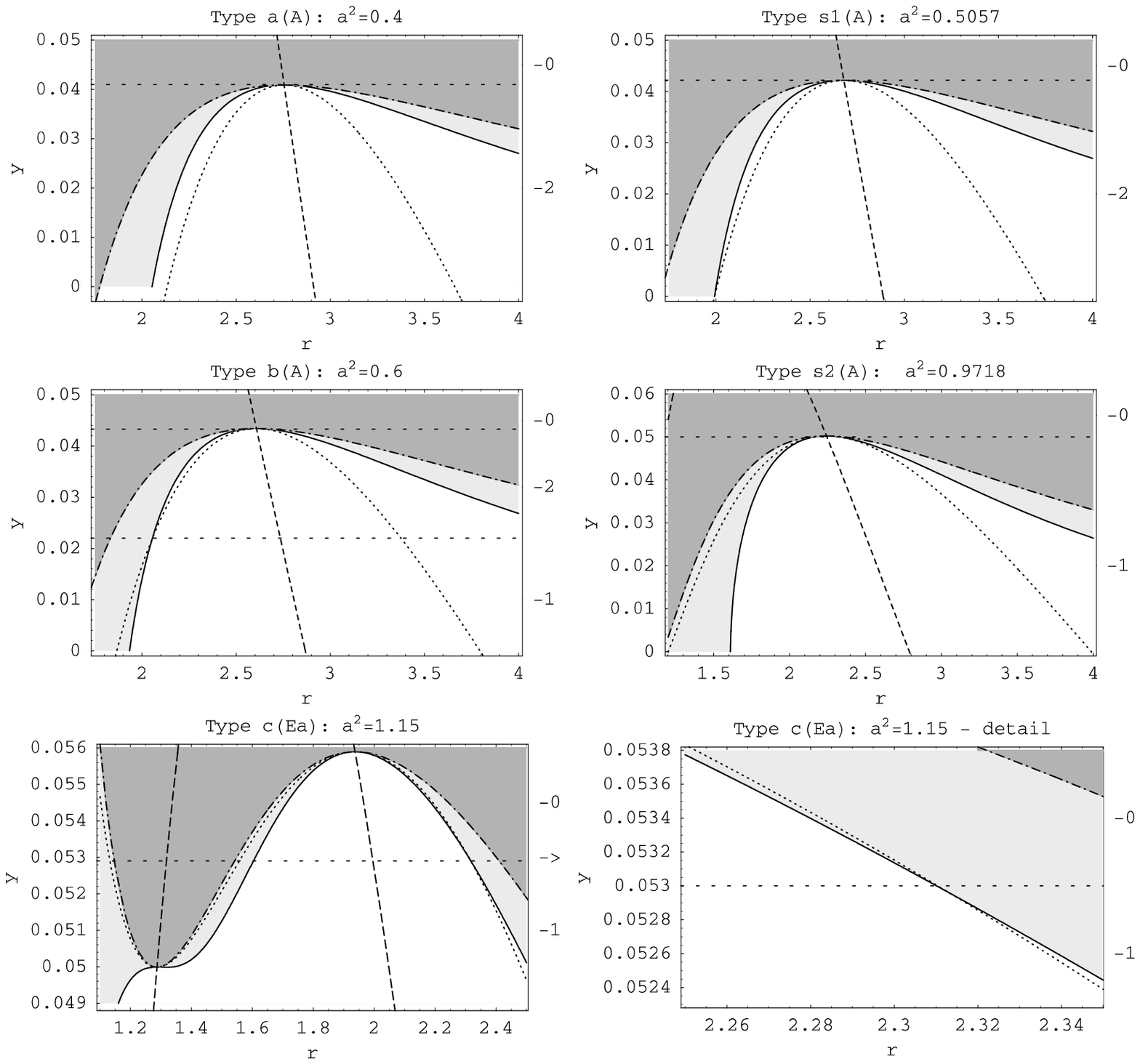}\end{center}
\caption{Functions $y_{ph}(r;a^2)$ (dotted) governing the radii of photon circular orbits; $y_{L\pm}(r;a^2)$ (solid) limiting embeddable regions (white); $y_T(r;a^2)$ (dashed) determining turning points of embedding diagrams; $y_h(r;a^2)$ (dashed-dotted) determining  locations of event horizons and limiting the dynamic regions (gray). In the parameter line ($y$) of the KdS spacetimes, the common points of $y_{ph}(r;a^2)$ and $y_{L\pm}(r;a^2)$ separate regions corresponding to different classes of the spacetimes differing in the number of embeddable photon circular orbits.}
\label{Fig:7} 
\end{figure}

\begin{figure}
\begin{center}\includegraphics[width=0.88\hsize]{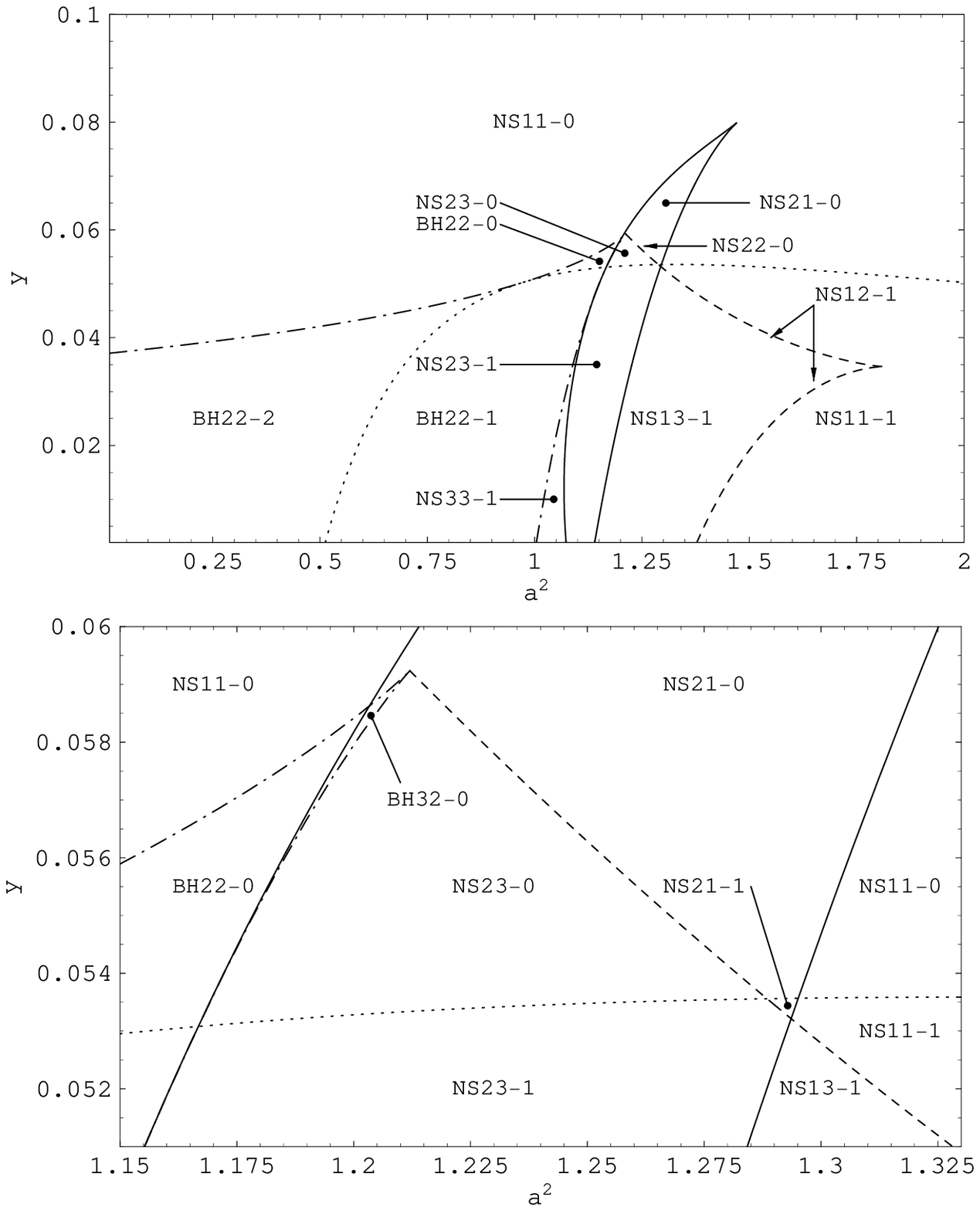}\end{center}
\caption{Classification of the KdS spacetimes according to the number of embeddable regions (first digit), turning points of embedding diagrams (second digit) and embeddable photon circular orbits (digit following dash). In the parameter plane $(a^2,y)$, the functions $y_{L+,e1}(a^2)$ and $y_{L+,e2}(a^2)$ (dashed-dotted) governing the extrema of the function $y_{L\pm}(r;a^2)$ coalescent with extrema of the function $y_{h}(r;a^2)$ (and $y_{ph}(r;a^2)$); $y_{L+,e3}(a^2)$, $y_{L+,e4}(a^2)$ and $y_{L-,max}(a^2)$ (solid) governing the remaining extrema of $y_{L\pm}(r;a^2)$; $y_{ph,L}(a^2)$ (dotted) governing the remaining common points of the functions $y_{ph}(r;a^2)$ and $y_{L\pm}(r;a^2)$; $y_{T,min}(a^2)$ and $y_{T,max}(a^2)$ (dashed) governing extrema of the function $y_{T}(r;a^2)$ separate regions corresponding to the different classes of the KdS spacetimes. Note that in the case of the classes NS12-1 and NS22-0, the second digits exceptionally denote one turning point and one inflexion point.} 
\label{Fig:8}
\end{figure}

Number of photon circular orbits depends on the number of extrema of the function $y_{ph}(r;a^2)$. Since these extrema coalesce with the extrema of the function $y_h(r;a^2)$ and there is $y_{ph}(r;a^2)\leq y_h(r;a^2)$ for all the values of $r>0$ and $a^2$, it clear that the KdS black-hole  spacetimes contain always two photon circular orbits in the outer stationary region and one orbit in the inner one. In the KdS naked-singularity spacetimes, only one such orbit always exists. However, all the orbits do not occur in the embeddable regions. The number of embeddable photon circular orbits depends on the number of common points of the functions $y_{ph}(r;a^2)$ and $y_{L\pm}(r;a^2)$ and their relative positions. We denote $y_{ph,L}(a^2)$ as the value of the function $y_{ph}(r;a^2)$ corresponding to the intersection of the functions  $y_{ph}(r;a^2)$ and $y_{L\pm}(r;a^2)$, determined by the function   
$a^2_{phL}(r)$, i.e., those intersections, which do not correspond to the extrema of the function  $y_{ph}(r;a^2)$. By using the characteristic functions $a^2_{phL}(r)$, $a^2_{he}(r)$ and $a^2_{ph0}(r)$, we can distinguish different types of the behaviour of the functions $y_{ph}(r;a^2)$ and $y_{L\pm}(r;a^2)$ (see figure~\ref{Fig:7} and table~\ref{Tab:3}). 
\begin{table}
\caption{\label{Tab:3} Different types of the KdS spacetimes determined by the corresponding different types of behaviour of the functions $y_{L\pm}(r;a^2)$ and $y_{ph}(r;a^2)$. The number of common points (except the common points coalescent with the extrema of $y_h(r;a^2)$) is expressed by the digits in related column, whereas the first digit denotes the number of common points 'between' the extrema of $y_{ph}(r;a^2)$ and the second one denotes the number of extrema 'behind' the maximum of $y_{ph}(r;a^2)$. Note that the types s1 and s2 represent limit cases of the behaviour of the functions. There is one negative common point of the functions in the case of the  type 'a',  denoted by the negative sign at the related digit.}
\begin{indented}
\item\begin{tabular}{@{}lll}
\br
\bf{T}&\bf{Range of} $\bf a^2$&\bf{Common points}\\ 
\hline
a&$(0,0.5057)$&-1,0\\ 
s1&$0.5057$&\;0,0\\ 
b&$(0.5057,0.9718)$&\;1,0\\ 
s2&$0.9068$&\;0,0\\ 
c&$(0.9068,\infty)$&\;0,1\\ 
\br
\end{tabular}
\end{indented}
\end{table}
In the parameter line ($y$) of the KdS spacetimes, the common points of $y_{ph}(r;a^2)$ and $y_{L\pm}(r;a^2)$ separate regions corresponding to different classes of the KdS spacetimes with two, one ore none embeddable photon circular orbits (see figure~\ref{Fig:7}).
The classification can be then obtained in the same way as the previous ones, i.e., by step by step discussing the number of embeddable circular photon orbits for each type of the spacetimes separately (see table~\ref{Tab:4}). The different classes of the KdS spacetimes are illustrated in figure~\ref{Fig:8}.
\begin{table}
\caption{\label{Tab:4} Classification of the KdS spacetimes according to the number of embeddable photon circular orbits.  
Limits of the ranges of the parameter $y$ (being functions of the parameter $a^2$) are illustrated in figure~\ref{Fig:8}.}
\begin{minipage}{0.38\linewidth}
\begin{indented}
\item\begin{tabular}{@{}ll}
\br
\bf{Class}&\bf{Range of} $\bf y$\\
\hline
{\it Type a}&\hspace{-2.5ex}\it{s1}\\
\hline
-2&$(0;y_{h,max}(a^2))$\\
-0&$\langle y_{h,max}(a^2);\infty)$\\
\hline
{\it Type b}&\\
\hline
-1&$(0;y_{phL}(a^2))$\\
-2&$\langle y_{phL}(a^2);y_{h,max}(a^2))$\\
-0&$\langle y_{h,max}(a^2);\infty)$\\
\br
\end{tabular}
\end{indented}
\end{minipage}
\hfill
\begin{minipage}{0.8\linewidth}
\begin{indented}
\item\begin{tabular}{@{}ll}
\br
\bf{Class}&\bf{Range of} $\bf y$\\
\hline
{\it Type s2}&\\
\hline
-1&$(0;y_{h,max}(a^2))$\\
-0&$\langle y_{h,max}(a^2);\infty)$\\
\hline
{\it Type c}&\\
\hline
-1&$(0;y_{phL}(a^2)\rangle$\\
-0&$(y_{phL}(a^2);\infty)$\\
\br\\
\end{tabular}
\end{indented}
\end{minipage}
\end{table}
%\begin{figure}
%\begin{center}
%\includegraphics[width=0.88\hsize]{figure11_revised1.eps}
%\end{center}
%\caption{Embedding diagrams of the optical reference geometry of the KdS spacetime of the classes %NS$21-0$ and NS$11-1$.}
%\label{Fig:11}
%\end{figure}
\section{Conclusions}
The presented study of the optical reference geometry and related inertial forces in the Kerr-de Sitter (KdS) spacetimes represents a generalization of results obtained earlier for the simpler limit cases of the KdS spacetimes, i.e., for the Kerr and Schwarzschild-de Sitter (SdS) spacetimes (see \cite{Stu-Hle:1999:ACTPS2:}, \cite{Stu:1990:BULAI:} and \cite{Stu-Hle:1999:PHYSR4:}). 

In the KdS spacetimes, effects of the rotation of source and the cosmic repulsion combine in a relatively complex way, being described by the rotational and cosmological parameters $a^2$ and $y$. In order to better understand the interplay of both the parameters, it is useful to summarize results of our investigation and compare them just with the studies on the background of the Kerr, SdS and Schwarzschild spacetimes, where the rotational and cosmological effects manifest separately.

The KdS black-hole spacetimes admit two stationary regions and two dynamic regions, separated by two black-hole horizons and one cosmological horizon. The KdS naked-singularity spacetimes then contain one stationary and one dynamic region separated by the cosmological horizon. 
Note that the optical reference geometry and related inertial forces are defined in the stationary (static) regions only.
The stationary and axially symmetric Kerr black-hole spacetimes ($a^2<1,\;y=0$) admit two stationary regions and one dynamic region, separated by two black-hole horizons. In the Kerr naked-singularity spacetimes ($a^2>1,\;y=0$), only one stationary region remains. 

The static and spherically symmetric SdS black-hole spacetimes ($a^2=0,\;0<y<1/27$) contain one static region, and two dynamic regions, separated by black-hole and cosmological horizons. 
In the SdS naked-singularity spacetimes ($a^2=0,\;y>1/27$), only one dynamic region exists. Therefore we do not pay attention to these naked-singularity spacetimes. The well known Schwarzschild black-hole spacetimes ($a^2=0,\;y=0$) contain the only horizon separating one dynamic region and one static region. 

An investigation of the optical geometry in a given spacetime can be considered divided into parts concerning the behaviour of the inertial forces, visualization of the optical geometry itself via embedding diagrams, and application of the formalism in relativistic dynamics. 

Of course, compared to the simpler cases of the Schwarzschild, SdS and Kerr spacetimes, the behaviour of the forces is more complex in the KdS spacetimes, incorporating the influence of both the spacetime parameters on the particle motion. There are more points of their vanishing, changing orientation and extrema. 
We have classified the KdS spacetimes according to the number of equatorial circular orbits where the velocity independent gravitational force vanishes and the orbits where the centrifugal force vanishes independently of the velocity (see table~\ref{Tab:5}). 
\begin{table}
\caption{\label{Tab:5} Classification of the KdS, Kerr, SdS and Schwarzschild spacetimes according to the number of circular orbits where the gravitational force vanishes (first digit), and orbits where the centrifugal force vanishes independently of the velocity (second digit).}
\begin{indented}
\item\begin{tabular}{@{}lll}
\br
&\bf{BH spacetimes}&\bf {NS spacetimes}\\ 
\hline
KdS&BH-22&NS-11, NS-21, NS-31, NS-32, NS-33\\ 
\hline
Kerr&BH-12&NS-01, NS-11, NS-21, NS-22, NS-23\\ 
\hline
SdS&BH-11&---\\ 
\hline
Schw.&BH-01&---\\ 
\br
\end{tabular}
\end{indented}
\end{table}
We can summarize that 
\begin{itemize}
\item KdS black-hole spacetimes contain only one circular orbit where the gravitational force vanishes, and only one orbit where the centrifugal force vanishes independently of the velocity in each of both the stationary regions. Both the gravitational and centrifugal forces change their orientations on these orbits. The same number of the orbits appears in the Kerr black-hole spacetimes, except for the outer stationary region, where there is no circular orbit where the gravitational force vanishes. 
In the SdS black-hole spacetimes, the gravitational force vanishes on the circular orbit of the static radius $r=y^{-1/3}$ only. The centrifugal force vanishes independently of the velocity (and $y$) on the orbit of the radius $r=3$. As well as in the Schwarzschild spacetimes, this is the only orbit where the centrifugal force vanishes independently of the velocity. On the other hand, the gravitational force does not vanish in the Schwarzschild spacetimes.   
\item KdS naked-singularity spacetimes can contain even three circular orbits where the gravitational force vanishes and other three orbits 
where the centrifugal force vanishes independently of the velocity, which indicates a relatively complex structure of these spacetimes as a result of mixed influence of the rotation of the source and the cosmological repulsion. The Kerr naked-singularity spacetimes contain at maximum only two circular orbits where the gravitational force vanishes, but there are at most three circular orbits where the centrifugal force vanishes independently of the velocity in accord with the case of the KdS naked-singularity spacetimes. With the parameters $a^2$ and $y$ growing, the structure of the KdS spacetimes simplifies. There are only one orbit with the vanishing of the centrifugal force and the other with the vanishing of the gravitational force.
\end{itemize}
The velocity independent part of the Coriolis force does not vanish neither in the KdS nor in the Kerr spacetimes, while in the SdS and Schwarzschild spacetimes the force does not occur at all.

In general, we can summarize, in static and spherically symmetric spacetimes, all the geodesics of the optical reference geometry coincide with trajectories of light, which explains the name 'optical geometry'. 
We have seen that test particles moving along the circular photon orbits are kept by the velocity-independent gravitational force only, because the remaining centrifugal force vanishes here independently of the velocity.      
Unfortunately, in the stationary  (non-static) spacetimes, the photon circular orbits in the equatorial plane do not correspond to the circular trajectories where the centrifugal force vanishes independently of the velocity. 

As for the embedding diagrams, the equatorial plane of the optical reference geometry can not be entirely embeddable into the 3D Euclidean space and the embedding diagrams then consist from several separated parts. 
We have characterized the shape of the embedding diagrams by the number of their turning points coalescing with the radii of circular orbits where the centrifugal force vanishes independently of the velocity. Because the radii of photon circular orbits in the equatorial are not located at the radii where the centrifugal force vanish, i.e., at the radii of the turning points of the diagrams, we have discussed embeddability of the photon circular orbits as well. 
Thus the KdS spacetimes can be also classified according to the features of the embedding diagrams (see table~\ref{Tab:6})).
\begin{table}
\caption{\label{Tab:6} Classification of the KdS, Kerr, SdS and Schwarzschild spacetimes according to the number of embeddable regions (first digit), turning points of the embedding diagrams (second digit) and the number of embeddable photon circular orbits (digit following dash).}
\begin{indented}
\item\begin{tabular}{@{}lll}
\br
&\bf{BH spacetimes}&\bf {NS spacetimes}\\ 
\hline
KdS&BH22-0, BH32-0&NS11-0, NS21-0, NS22-0, NS23-0, NS11-1, NS12-1\\ 
   &BH22-1, BH22-2&NS13-1, NS21-1, NS22-1, NS23-1, NS33-1\\ 
\hline
Kerr&BH22-2, BH22-1&NS11-1, NS12-1, NS13-1, NS23-1, NS33-1\\ 
\hline
SdS&BH11-1&---\\ 
\hline
Schw.&BH11-1&---\\
\br
\end{tabular}
\end{indented}
\end{table}
We can summarize that 
\begin{itemize}
\item KdS black-hole spacetimes contain three or two separated embeddable regions in contrast to the Kerr black-hole spacetimes, where only two regions occur, and to the SdS and Schwarzschild black-hole spacetimes with the only embeddable region. 
\item KdS naked-singularity spacetimes contain one, two or three separated embeddable regions as well as the Kerr naked-singularity spacetimes. 
%***********************************************************88
\item KdS black-hole embedding diagrams contain two turning points as well as the Kerr black-hole diagrams. But the SdS and Schwarzschild black-hole embedding diagrams contain one turning point, coalescing with the radius of the circular orbit where the centrifugal force vanishes independently of the velocity, i.e., with the radius $r=3$.
\item KdS naked-singularity embedding diagrams contain one, two or three turning points as well as  the Kerr naked-singularity diagrams.
%**********************************************************88
\item KdS black-hole spacetimes contain three photon circular orbits, whereas two of them occur in the outer stationary region and one of them in the inner one, as well as in the Kerr black-hole spacetimes, but in contrast to the SdS and Schwarzschild black-hole spacetimes, where only one photon orbit occurs at the radius $r=3$.
\item KdS naked-singularity spacetimes contain one photon circular orbit as well as the Kerr naked-singularity spacetimes. 
%*************************************************************
\item KdS black-hole spacetimes contain two, one or none embeddable photon circular orbits in the outer stationary region in contrast to the Kerr black-hole spacetimes, where always two or one embeddable orbits exist, and to the SdS and Schwarzschild black-hole spacetimes, where always one embeddable orbit occurs at the radius coalescing with the radius of turning point of the embedding diagram. 
\item KdS naked-singularity spacetimes contain one or none embeddable photon circular orbit in contrast to the Kerr naked-singularity spacetimes, where always one embeddable orbit occurs.
\end{itemize}\quad\\\\
%-----------------------------------
Both the presented parts of the investigation of optical reference geometry formalism and related classification of the KdS spacetimes can be consider to be the basic and necessary study,  before the formalism is applied in the dynamics.

By using some particular examples of the particle and fluid orbital motion, we have just directly demonstrated that the optical reference geometry and related inertial forces formalism presents certain, we can say 'pseudo-Newtonian' approach in relativistic dynamics. This approach is based on the Newtonian way of dynamics description, but taking the general relativistic effects fully into account. 
Now, we can state that the simplicity and applicability of the formalism, which were only predicted in a general way in some papers (e.g. \cite{Abr-Nur-Wex:1995:CLAQG:}) or directly shown, but only for the simpler types of spacetimes, 'survive' even in the complex types of the KdS spacetimes, where its application is then more useful all the more just because of the complexity of the spacetimes.
     
In spite of the more or less technical aspect of the paper, the study can be reviewed from the astrophysical point of view as well. We have shown that the combined effect of rotation and cosmic 
repulsion is relevant in both the cases of black-hole and naked-singularity backgrounds. 
Clearly, choosing a characteristic type of motion and knowing the related behaviour of the forces, we can understand the influence of the rotational and cosmic parameters. Of course, there are many astrophysical and conceptual reasons to choose the circular motion, as we have done. 
Among others, comparing the results for the Kerr and KdS spacetimes, the additional radius in the KdS spacetimes, where the gravitational force vanishes, the same number of circular orbits where the centrifugal force vanishes and qualitatively same types of behaviour of embedding diagrams suggest that the influence of the cosmological constant is fundamentally imprinted mainly in the gravitational force. Note that it was predicted in \cite{Stu:1990:BULAI:}, by comparing the behaviour of the gravitational force in the SdS and Schwarzschild spacetimes (see tables~\ref{Tab:5} and \ref{Tab:6}).

One of the important new results concerns properties of the so-called static radii, where the gravitational attraction of a black hole (naked singularity) is just balanced by the cosmic repulsion. We have found that in the equatorial plane the location of the static radius $r=y^{-1/3}$ is independent of the rotational parameter $a$, and formally coincides with the results from the SdS black-hole spacetimes, while on the axis of symmetry it depends on $a$. Physically, it is important that these results are easily implied by the analysis of the behaviour of the inertial forces.

From the physical point of view, the most important new result is given by the possibility to derive the Euler equation easily in the framework of optical geometry and related inertial forces as shown in Section 3. It enables us to treat behaviour of perfect fluid orbiting black holes, and in general all the accretion processes, in the pseudo-Newtonian way related to the inertial forces. We believe the pseudo-Newtonian approach could be useful in modelling self-graviting discs around SdS and KdS black-holes. It should be noticed in this connection that character of the centrifugal force, which is relevant in determining the structure of thick discs, is clearly reflected by the embedding diagrams of the optical geometry (see \cite{Stu-Hle-Jur:2000:CLAQG:}).   

Thus we can state, we dispose an alternative method to the standard general relativistic approach, practically and intuitively applicable in the KdS spacetimes, and hence suitable for the description of astrophysical phenomena around supermassive black holes in centre of giant active galactic nuclei.
%----------------------------------------------------------  
%******************************************
\ack
This work was supported by the Czech grant MSM 4781305903.
%***********************************************************************

%*******************************************************************************
\end{document}